\documentclass[manuscript,screen]{acmart}
\AtBeginDocument{%
  }

\usepackage{multirow}      % For multi-row cells
\usepackage{xcolor}        % For colors (load BEFORE colortbl)
\usepackage{colortbl}      % For row coloring
\usepackage{amsmath}       % For checkmark symbol

\usepackage{amssymb}       % For additional symbols

\usepackage{dblfloatfix}

\usepackage{algorithm}
\usepackage{algpseudocode}
\usepackage{booktabs}
\usepackage{hyperref}

\usepackage{placeins}

\newcommand{\cmark}{\checkmark}
\newcommand{\xmark}{\texttimes}

\usepackage{tikz}
\usetikzlibrary{trees}
\usetikzlibrary{shapes,arrows,positioning,fit,calc}

\setcopyright{none}
\begin{document}

\title{FARM: Field-Aware Resolution Model for Intelligent Trigger-Action Automation}

%% ORCID is REQUIRED for all TIST authors
\author{Khusrav Badalov}
\orcid{https://orcid.org/0009-0009-6906-919X}  % Replace with your actual ORCID
\affiliation{%
  \institution{Neouly Co., Ltd.}
  % \department{Computer Engineering}
  \city{Seoul}
  \country{Republic of Korea }
}
\email{khusravvvb99@gmail.com}

\author{Young Yoon}
\orcid{https://orcid.org/0000-0002-5249-2823}  % Replace with your actual ORCID
\affiliation{%
  \institution{Hongik University}
  \department{Computer Engineering}
  \city{Seoul}
  \country{Republic of Korea}
}
\email{young.yoon@hongik.ac.kr}

%% Set shorter author list for headers if needed
\renewcommand{\shortauthors}{Badalov and Yoon}





\begin{abstract}
Trigger-Action Programming (TAP) platforms such as IFTTT and Zapier enable Web of Things (WoT) automation by composing event-driven rules across heterogeneous services. A TAP applet links a \textit{trigger} to an \textit{action} and must bind trigger outputs (\textit{ingredients}) to action inputs (\textit{fields}) to be executable. Prior work largely treats TAP as service-level prediction from natural language, which often yields non-executable applets that still require manual configuration. We study the \textit{function-level configuration} problem: generating complete applets with correct ingredient-to-field bindings. We propose FARM (Field-Aware Resolution Model), a two-stage architecture for automated applet generation with full configuration. Stage~1 trains contrastive dual encoders with selective layer freezing over schema-enriched representations, retrieving candidates from 1,724 trigger functions and 1,287 action functions (2.2M possible trigger-action pairs). Stage~2 performs selection and configuration using an LLM-based multi-agent pipeline. It includes intent analysis, trigger selection, action selection via cross-schema scoring, and configuration verification. Agents coordinate through shared state and agreement-based selection. FARM achieves 81\% joint accuracy on Gold (62\% Noisy, 70\% One-shot) at the function level, where both trigger and action \emph{functions} must match the ground truth. For comparison with service-level baselines, we map functions to their parent services and evaluate at the service level. FARM reaches 79\% joint accuracy and improves over TARGE by 21 percentage points. FARM also generates ingredient-to-field bindings, producing executable automation configurations.

\end{abstract}

\maketitle

%%%%%%%%%%%%%%%%%%%%%%%%%%%%%%%%%%%%%%%%%%%%%%%%%%%%%%%%%%%%%%%%%%%%%%%%%%%%%
\section{Introduction}\label{sec:introduction}

Trigger-Action Programming (TAP) is a declarative programming paradigm that allows users to define event-driven automation rules. It is the essential foundation behind systems such as IFTTT (If This Then That), Zapier, and similar platforms including Microsoft Power Automate and Apple Shortcuts. Ur et al.\ \cite{ur2014practical} demonstrated the practicality of TAP in smart home contexts, while Rahmati et al.\ \cite{rahmati2017ifttt} provided a comparative analysis of IFTTT and Zapier. TAP has become a mainstream interface for end-user automation across Web of Things (WoT) ecosystems, connecting both IoT devices and web applications; Mi et al.\ \cite{mi2017ifttt} found that over 400 services spanning both categories are integrated on IFTTT alone. Ur et al.\ \cite{ur2016tapwild} showed that real-world TAP rule corpora exhibit substantial diversity in services, rule structures, and user-written natural language descriptions, making TAP a challenging target for robust intent understanding and program synthesis.

\begin{figure}[h]
  \centering
  \includegraphics[width=1.0\linewidth]{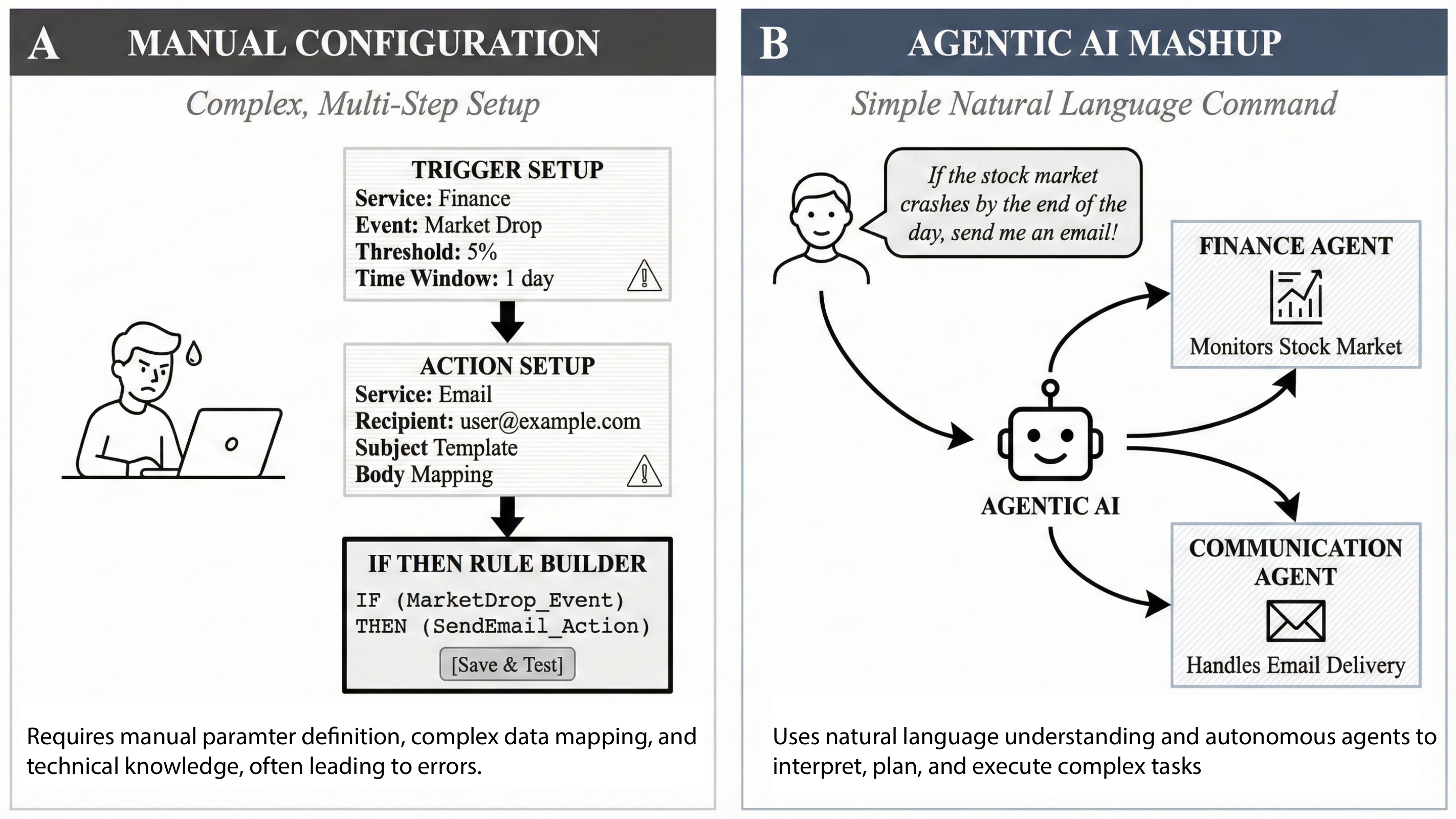}
  \caption{An example of an TAP applet interface for manual configuration and our proposal for an automated applet generation.}
  \label{fig:ifttt-example}
\end{figure}

Historically, TAP interaction was associated with composing simple IF-THEN rules by manually selecting a trigger and an action and combining them into a mashup \cite{huang2015mental,rahmati2017ifttt}. Recently, LLM-driven assistants and agentic systems have shifted this workflow toward natural language specifications, where a model is expected to (i) infer user intent, (ii) select services, and (iii) produce executable configurations \cite{king2023sasha,shi2024awareauto}. Figure~\ref{fig:ifttt-example} illustrates this paradigm shift, contrasting traditional manual configuration with agentic AI-driven mashup generation. However, in practice, AI-based TAP builders frequently fail on underspecified or ambiguous requests: users often phrase the same intent in multiple ways, provide incomplete context (e.g., location, device identity, thresholds), or implicitly encode preferences (e.g., comfort vs.\ energy saving) \cite{liu2023ambiguity,kim2024iotgpt}. In such cases, current systems may generate multiple inconsistent trigger-action variants, select incorrect parameters, or produce brittle applets that are syntactically valid but semantically misaligned with user goals.

%%%%%%%%%%%%%%%%%%%%%%%%%%%%%%%%%%%%%%%%%%%%%%%%%%%%%%%%%%%%%%%%%%%%%%%%%%%%%
\subsection{Motivation}
Consider a common scenario: Alice asks an AI assistant, ``Notify me when it's going to rain tomorrow.'' Although seemingly straightforward, existing TAP platforms frequently mishandle such requests: alerts may trigger for the wrong location, apply an unintended threshold, or fail due to missing implicit parameters. This example highlights a central challenge in contemporary TAP systems: the widening semantic gap between informal user language and the structured, parameter-rich function-level invocations required for reliable automation. 
% ADDED: Operation-level definition
We use the term \emph{function-level} to distinguish from service-level identification: while a service (e.g., ``Google Sheets'') may expose multiple triggers and actions, a function refers to a specific trigger (e.g., ``New row added to spreadsheet'') or action (e.g., ``Update cell in spreadsheet''), each with its own input parameters, output data fields, and schema constraints.
Prior work also shows that users struggle to debug such failures, because the root cause may lie in hidden parameters, platform semantics, or multi-rule interactions Zhang et al. \cite{zhang2023debugtap}.

Recent advances in TAP generation have made significant strides. Yusuf et al.\cite{yusuf2022} demonstrated that transformer seq2seq models can generate TAPs with improved service and field accuracy from natural language descriptions. Liu et al. \cite{liu2016latent} learning-based approaches, including attention and latent alignment formulations, framed TAP inference as service/action prediction under weak supervision . However, these state-of-the-art works largely focus on predicting the correct services and fields, and do not explicitly ensure that generated configurations are aligned with the user's higher-level intent or preferences.

Conversely, Wu et al.\ \cite{wu2025mvtap} pioneered implicit intention prediction through multi-view representation learning in their MvTAP framework, showing that users often create rules driven by latent goals (e.g., ``energy saving'' when turning off devices). Yet MvTAP addresses intention classification rather than the end-to-end generation of executable applet configurations with function-level parameters. In parallel, Hsu et al.\ \cite{hsu2019safechain} showed that even seemingly correct TAP rules can yield undesirable outcomes due to cross-rule interactions and hidden automation chains, reinforcing that correctness must be defined at the level of executable behavior, not only service accuracy.

This dichotomy between generation accuracy and intention understanding represents a critical gap. A system can generate syntactically valid TAPs while still missing the user's actual intent. On the other hand, a system can correctly infer, for instance, ``energy saving'', without being able to translate that goal into concrete trigger conditions and action parameters. In addition, the move toward agentic LLM tool use introduces a further constraint: systems must reliably decide \emph{when} to call tools, \emph{which} tools to call, and \emph{how} to parameterize calls, under uncertainty in the user request Schick et al.\cite{schick2023toolformer}.

To clearly illustrate these challenges, we performed a systematic study on IFTTT and identified recurrent failure patterns that neither intent-only nor generation-only approaches fully address:

\begin{itemize}
    \item \textbf{Understanding the Core of Services:} Services vary widely in functional richness. Popular Web services such as Google Drive, X (formerly Twitter), and Spotify expose many triggers and actions, while WoT device services such as smart thermostats or light sensors are often sparse due to hardware constraints. This induces severe imbalance in training data and increases the risk of overfitting to richly represented services \cite{ur2016tapwild}.

    \item \textbf{Intention-Structure Mismatch:} Classification-based approaches can identify latent intent, and generation-based approaches can produce valid structures. However, aligning intent with structural choices (trigger conditions, action parameters, and field values) remains unreliable \cite{wu2025mvtap,yusuf2022}.

    \item \textbf{Context-Dependent Field Selection:} Field names alone are insufficient. The same services may require different threshold values or options depending on intent (e.g., comfort-focused vs.\ energy-saving automations), and current generators are weak at selecting values that reflect user goals \cite{zhang2023debugtap}.

    \item \textbf{Cross-Functional Reasoning:} Similar intents can couple disparate services (e.g., weather plus messaging, occupancy plus HVAC). Existing approaches often treat service selection as independent, missing intent-driven cross-service dependencies \cite{wu2025mvtap}.
\end{itemize}

These observations motivate our core insight: effective TAP configuration requires simultaneously understanding user intentions and generating accurate technical implementations. Traditional approaches, whether intent-centric or generation-centric, address only one side of this dual requirement.

Through experimentation and analysis of 16k+ real IFTTT applets with enriched metadata, we find that neither intention understanding nor  structure generation alone suffices. The remaining challenge is architectural: bridging intent comprehension with executable configuration generation. This motivates our multi-agent approach that decomposes the problem and coordinates complementary capabilities.

%%%%%%%%%%%%%%%%%%%%%%%%%%%%%%%%%%%%%%%%%%%%%%%%%%%%%%%%%%%%%%%%%%%%%%%%%%%%%
\subsection{Research Questions}
Our investigation was guided by the following research questions:

\begin{enumerate}
    \item \textbf{How can we handle severe service-level imbalance across TAP domains, where some platforms expose dozens of triggers and actions while others provide only a few?}

    The TAP ecosystem exhibits extreme distributional skew: popular services like Google Drive, X (formerly Twitter), or Spotify have rich function-level coverage, while niche WoT devices such as Philips Hue or Nest Thermostat may expose only one or two triggers. This imbalance leads to underrepresentation of sparse function-level metadata in learned representations, causing models to favor well-represented services regardless of user intent \cite{ur2016tapwild}. We investigate whether training separate encoders for triggers and actions---combined with a layer freezing strategy that preserves pretrained semantic knowledge---can achieve reliable retrieval across both common and rare services.

    \item \textbf{How can a multi-agent architecture decompose the TAP generation task to produce complete, executable applet configurations rather than just service name predictions?}

    Prior approaches predict only which services to use, leaving users to manually configure fields and parameters. We investigate whether separating the problem into specialized agents---one for understanding trigger events, another for reasoning about action requirements, and a third for verifying compatibility---can generate complete configurations including field bindings that map trigger outputs to action inputs.
\end{enumerate}

%%%%%%%%%%%%%%%%%%%%%%%%%%%%%%%%%%%%%%%%%%%%%%%%%%%%%%%%%%%%%%%%%%%%%%%%%%%%%
\subsection{Contributions}
To summarize, this work makes the following contributions:

\begin{itemize}
    \item We introduce \textbf{FARM}\footnote{Code: \url{https://github.com/DinaHongik/FARM}. Dataset available upon request: \texttt{young.yoon@hongik.ac.kr}} (Field-Aware Resolution Model), a two-stage framework that combines contrastive-trained dual encoders for high-recall candidate retrieval with multi-agent LLM-based selection for precise configuration generation.
    
    \item We propose a \textbf{layer freezing strategy} for domain-specific encoder fine-tuning that preserves pretrained semantic knowledge while adapting to trigger-action retrieval, achieving over 90\% recall at rank 5 for both triggers and actions.

    \item We design a \textbf{four-agent selection pipeline} where specialized agents (Intent Analyzer, Trigger Selector, Action Selector, Verifier) coordinate to generate complete function-level configurations with field bindings---not just service names.
    % ADDED: Clarification of what operation-level configuration includes
    These configurations include trigger input parameters, action required fields, and ingredient-to-field mappings that specify how trigger outputs populate action inputs.

    \item We demonstrate that FARM achieves \textbf{79\% joint accuracy} on clear queries, outperforming prior approaches by +21 points over TARGE \cite{cimino2025targe}, while additionally generating executable applet configurations that prior methods do not produce.

    \item We evaluate FARM across three test conditions---clear queries (Gold data), ambiguous queries (Noisy data), and rare function-level queries (One-shot data) ---demonstrating robust performance even on challenging inputs where existing methods struggle.
\end{itemize}

\section{Background and Related Work}\label{sec:background}

Many existence methods have been conducted in the area of trigger-action programming (TAP), and we divided most of the relevant research into two groups: 1) Rule synthesis and generation 2) Rule searching and classification.

Early efforts on automatic trigger-action program (TAP) composition treated the problem as a multi-class classification task mapping natural language (NL) descriptions to pre-defined trigger-action functions. Quirk et al.~\cite{quirk2015} first framed TAP generation as classification by training a binary logistic regression classifier for each possible "if-then" rule component. Their system extracted linguistic features (unigrams, bigrams, character trigrams) from the NL description and predicted which trigger or action functions should appear in the resulting recipe. While effective, this approach required one classifier per candidate function, which did not scale as new IoT services and functions emerged. Subsequent classification-based approaches improved efficiency and accuracy. Yoon et al.~\cite{yoon2016} has been proposed a CRF-based learning method that identifies relevant trigger services and predicts trigger-action pairs for user requests. This approach combined information retrieval with parallel learning engines, outperforming earlier single-model classifiers in both accuracy and training time. Deep learning further enhanced classification methods: the Latent Attention Model (LAM) by Liu et al.~\cite{liu2016latent} trained separate neural classifiers for triggers and actions however introduced a "latent" attention mechanism (LAM) to weight words in the description by importance. LAM achieved state-of-the-art accuracy at the time, yet still treated trigger and action prediction independently, failing to capture their inter dependencies. In general, pure classification approaches are fast and straightforward but tend to struggle when user descriptions are implicit or vague, since they ignore the joint semantics of triggers and actions. To address the limitations of disjoint classification, researchers turned to generative or semantic parsing techniques that model TAP creation as a structured prediction or sequence-to-sequence generation problem. Beltagy and Quirk~\cite{beltagy2016improved} recast TAP synthesis as constructing a parse tree (sequence of rule components) rather than independent classifications. Their system predicted one component of the recipe at a time, conditioned on the NL input and previously generated components, using an ensemble of logistic regression and a multilayer perceptron. This structured approach outperformed the earlier purely binary classification by Quirk et al.~\cite{quirk2015} and mitigated the scalability issue, although it still relied on bag-of-words text features and thus ignored deeper semantic context. More advanced generative models leverage modern neural sequence learning. Yao et al.~\cite{yao2019interactive} developed an interactive semantic parsing framework where a conversational agent learns to synthesize if-then rules through dialogue, employing hierarchical reinforcement learning to decide on sub-goals and rule components. This interactive approach allowed the agent to ask clarifying questions and handle complex recipes, but training it required costly simulation of user interactions. In contrast, one-shot generation models aim to produce the rule in a single pass. Yusuf et al.~\cite{yusuf2022} introduced RecipeGen, a Transformer-based sequence-to-sequence model that directly "translates" a NL description into a trigger-action script. By framing TAP creation as language generation rather than classification, their approach could learn the implicit relationships between trigger and action phrases (e.g., understanding that "photo tagged with me on Facebook" implies a Facebook file URL trigger and a Dropbox upload action). To boost accuracy, RecipeGen was warm-started with a pre-trained encoder, adapting it to the IFTTT domain vocabulary. This generative model significantly outperformed the LAM classifier on multiple real-world TAP datasets, achieving higher recall and BLEU scores for correct trigger-action predictions. The success of such sequence-learning methods demonstrates that modeling the joint structure of TAPs can handle unclear or complex user requests better than independent classifiers. However, generative models must ensure the validity of produced rules (respecting available services/fields), often addressed through constrained decoding or post-validation.

A third line of work uses hybrid approaches that combine data-driven learning with domain knowledge or that recast TAP creation as a recommendation problem. Instead of parsing free-form descriptions, these methods often assume partial user input (such as chosen trigger or a high-level goal) and leverage existing rule repositories or ontologies to suggest the rest of the rule. Corno et al.~\cite{corno2019recrules} developed RecRules, a recommendation system that represents IoT channels and functions in a semantic graph and uses collaborative filtering on past user-created rules. Given a user's current context or usage history, RecRules performs semantic reasoning over the graph (using manually crafted ontological rules) and then suggests likely trigger-action combinations based on similarity to what other users have done. This knowledge-driven system improved the relevance of suggestions but required maintaining an expert-defined semantic model of the domain.

To incorporate explicit user intentions in automation, Corno et al.~\cite{corno2020taprec} proposed TAPrec, which introduced an ontology (EUPont) of goal-oriented activities to guide rule creation. In TAPrec, when a user selects a trigger, the system infers the user's high-level goal (e.g. "personalize room lighting") by mapping the trigger to an OWL class in the ontology, and then recommends an action that fulfills that goal. This approach effectively casts action selection as a goal-conditioned classification task, constrained by expert knowledge of typical IoT goals. They later extended this idea with a conversational assistant called HeyTAP~\cite{corno2021from} which asks the user questions in natural language, extracts an abstract intention from the dialogue, and matches it to predefined goal classes to recommend an appropriate if-then rule. These ontology-backed systems emphasize semantic constraints (ensuring the recommended trigger and action logically serve the same goal), but they come with the overhead of manually constructing and updating the ontology and labeling rules with intention metadata. Recent research strives to reduce this manual effort by using learning on graph-structured representations. Huang et al.~\cite{huang2023tap} introduced TAP-AHGNN, an attention-based heterogeneous graph neural network that learns embeddings for TAP components (devices, triggers, actions) and uses them to recommend services. TAP-AHGNN integrates a knowledge graph of the IoT domain with GNNs, automatically capturing relationships (e.g. which devices or services are often used together) and predicting the next action given a trigger context. This hybrid model leverages structured knowledge and data-driven patterns, outperforming purely collaborative or content-based recommenders by learning complex cross-service associations (while still requiring an initial knowledge graph schema). Another emerging trend is applying large language models: King et al.~\cite{king2023sasha} present Sasha, a smart-home assistant that feeds ambiguous voice commands and home context into an LLM to generate possible automation rules. Sasha showed the viability of general-purpose LLMs for TAP recommendation, but still needed each training rule to be labeled with an intention (goal) to guide the model.

\subsection{Comparison and Limitations of Previous Works}
\subsubsection{Comparison}
The evolution from classification-based to generative and hybrid approaches demonstrates a steady increase in the capability of TAP automation systems. Early classification methods (e.g., Quirk et al.~\cite{quirk2015}, Yoon et al.~\cite{yoon2016}, Kuang et al.~\cite{Kuang2025SAFE-TAP}) were effective for known trigger–action pairs but faced challenges in scalability and capturing semantic nuances.
Generative approaches (e.g., Yusuf et al.~\cite{yusuf2022}, Yao et al.~\cite{yao2019interactive}, Liu et al.~\cite{liu2021tagen}, Liu et al.~\cite{liu2023generating}), ranging from structured predictors to neural sequence-to-sequence models, introduced greater flexibility by jointly modeling triggers and actions while using context from natural language descriptions. These methods achieved improved accuracy on complex user requests and unseen combinations.

\subsubsection{Limitations of Previous Work}

\begin{figure*}[htbp]
    \centering
    \includegraphics[width=0.80\textwidth]{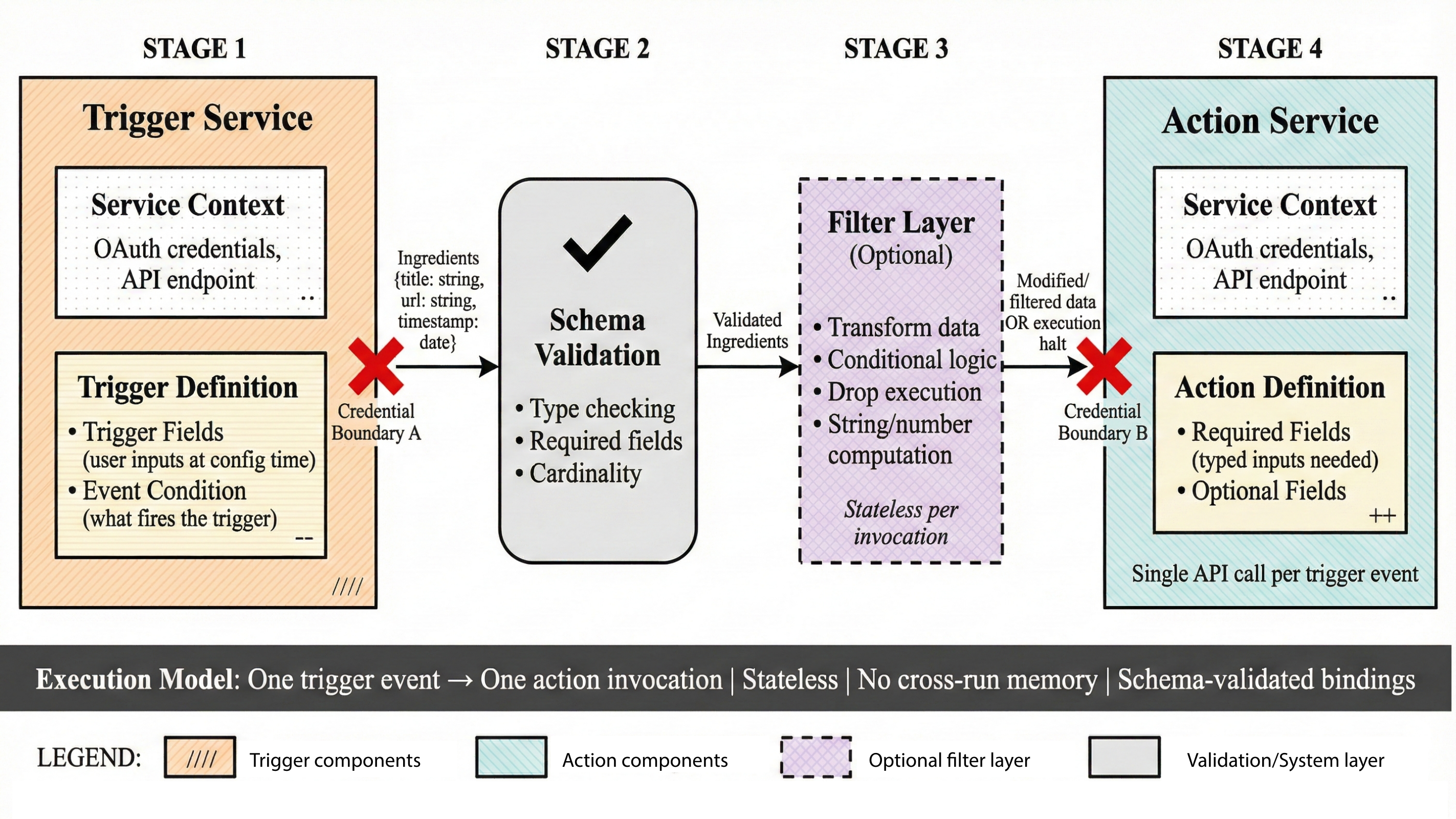}
    \caption{IFTTT applet execution pipeline with credential boundaries.}
    \label{fig:ifttt-execution}
\end{figure*}

While the methods mentioned above improve the efficiency and accuracy of user-created TAP rules, to the best of our knowledge, none of the existing TAP research explains how a user's natural language query can directly specify which function-level parameters and fields are required, nor how these can be used to automatically mash up two independent function—trigger and action—into a complete applet. As illustrated in Fig.~\ref{fig:ifttt-execution}, IFTTT applets operate across \emph{credential boundaries} (marked with $\times$) that isolate each service's authentication context---the trigger service's OAuth credentials never reach the action service, and vice versa. This security model means that only validated data (ingredients) flows between stages, requiring any automated system to understand both the schema validation requirements and the field-level mappings needed to bridge these isolated services.

Beyond this configuration gap, existing approaches suffer from limited expressivity in handling complex automation scenarios. The simple one-to-one trigger-action mapping requires an "impedance match" between trigger outputs and action inputs at a shared abstraction level. When deeper customization is needed—such as combining data from multiple sources or applying conditional transformations—users must write external "glue code" or resort to traditional programming methods. This limitation prevents platforms from supporting real-world scenarios involving multiple conditions, sequential triggers, or stateful computations, ultimately restricting the practical utility of automated TAP generation systems.
Furthermore, current generation methods produce rules in isolation without considering inter-rule interactions or security implications. When multiple TAP rules are simultaneously enabled, complex system behaviors emerge that are difficult to diagnose. Wang et al.~\cite{wang2019charting} demonstrated that 66\% of synthetic IFTTT deployments exhibit potential inter-rule vulnerabilities, including action conflicts and infinite actuation loops. Existing TAP synthesis approaches focus solely on functional correctness for individual rules, lacking mechanisms to verify compatibility with a user's existing rule set or to ensure that generated configurations do not introduce unintended security risks.
Table~\ref{tab:tap_comparison} provides a comprehensive comparison of existing approaches.

\begin{table*}[!t]
    \centering
    \caption{Comparison of TAP Automation Approaches: Functional and Non-Functional Aspects}
    \label{tab:tap_comparison}
    \resizebox{\textwidth}{!}{%
    \begin{tabular}{@{}llllccc|ccc@{}}
    \toprule
    \multirow{2}{*}{\textbf{Authors}} &
    \multirow{2}{*}{\textbf{Method}} &
    \multirow{2}{*}{\textbf{Input}} &
    \multirow{2}{*}{\textbf{Output}} &
    \multicolumn{3}{c|}{\textbf{Functional Capabilities}} &
    \multicolumn{3}{c}{\textbf{Non-Functional Aspects}} \\
    \cmidrule(lr){5-7} \cmidrule(lr){8-10}
    & & & &
    \textbf{Compos.} &
    \textbf{Iterative} &
    \textbf{Service} &
    \textbf{Reported} &
    \textbf{Dataset} &
    \textbf{Data} \\
    & & & &
    \textbf{Reasoning} &
    \textbf{Refine} &
    \textbf{Extend.} &
    \textbf{Accuracy} &
    \textbf{Scale} &
    \textbf{Require.} \\
    \midrule
    % Classification Methods
    Quirk et al. \cite{quirk2015} &
    Binary &
    NL query &
    Executable &
    \xmark &
    \xmark &
    \xmark &
    Channel: 50\% &
    114K recipes &
    High \\
    & Classification & & AST & (independent) & & (retrain) & Function: 37\% & 160 channels & (labeled pairs) \\
    \midrule
    Yoon et al. \cite{yoon2016} &
    CRF-based &
    NL query &
    Service &
    \xmark &
    \xmark &
    \xmark &
    MA: 51.6\% &
    270K pages &
    Medium-High \\
    & Learning & & names & (independent) & & (retrain) & NE: 32.7\% & 1K-9K train & (seq. labels) \\
    \midrule
    Liu et al. \cite{liu2016latent} &
    Neural &
    NL query &
    Service &
    \xmark &
    \xmark &
    \xmark &
    87.5\% &
    68K train &
    High \\
    (LAM) & Attention & & names & (independent) & & (retrain) & (joint pred.) & 584 test & (labeled pairs) \\
    \midrule
    % Generation Methods
    Beltagy \& Quirk \cite{beltagy2016improved} &
    Structured &
    NL query &
    Executable &
    Partial &
    \xmark &
    \xmark &
    Tree: 42.6\% &
    77.5K train &
    High \\
    & Prediction & & JSON & (sequential) & (1-shot) & (retrain) & Channel: 54\% & 5.2K dev, 4.3K test & (labeled deriv.) \\
    \midrule
    Yusuf et al. \cite{yusuf2022} &
    Transformer &
    NL query &
    Executable &
    Implicit &
    \xmark &
    \xmark &
    BLEU: 59.1\% &
    45K-120K &
    High \\
    (RecipeGen) & Seq2Seq & & Output & (decoder) & (1-shot) & (retrain) & SR@1: 50.6\% & merged sets & (labeled pairs) \\
    & & & & & & & MRR@3: 0.575 & & \\
    \midrule
    Yao et al. \cite{yao2019interactive} &
    Hierarchical &
    Dialogue &
    Executable &
    \cmark &
    \cmark &
    \xmark &
    Sim: 89.4\% &
    291K recipes &
    High \\
    & RL & & JSON & & (dialogue) & (retrain) & Human: 63.4\% & + sim. dialogue & (recipes + sim.) \\
    \midrule
    % Recommendation Methods
    Corno et al. \cite{corno2020taprec} &
    Ontology &
    Partial &
    Service &
    \cmark &
    \xmark &
    Manual &
    N/A &
    Ur et al. scale &
    Medium \\
    (TAPrec) & Reasoning & selection & names & (goal-based) & & update & (recommends) & + EUPont ont. & (rules + ont.) \\
    \midrule
    Huang et al. \cite{huang2023tap} &
    GNN &
    Context &
    Service &
    Partial &
    \xmark &
    Partial &
    HR@10: 0.937 &
    9,884 recipes &
    Medium \\
    (TAP-AHGNN) & Embeddings & & names & & & (graph) & NDCG@10: 0.952 & 1,249 triggers & (graph struct.) \\
    & & & & & & & MRR@10: 0.967 & 748 actions & \\
    \midrule
    Wu et al. \cite{wu2025mvtap} &
    Multi-view &
    Multi-view &
    User &
    \xmark &
    \xmark &
    \xmark &
    Micro-F1: 0.783 &
    N/A &
    Medium \\
    (MvTAP) & Learning & (U/D/K) & intentions & (independent) & & (retrain) & Macro-F1: 0.751 & N/A & (multi-label) \\
    \midrule
    % TARGE
    Cimino et al. \cite{cimino2025targe} &
    Cross-view &
    NL query &
    Executable &
    \cmark &
    \xmark &
    Partial &
    EM: 61\% (gold) &
    34.8K rules &
    Medium-High \\
    (TARGE) & Contrastive & (user intent) & rules & (CRG: trigger- & (1-shot, & (clustering, & EM: 41\% (noisy) & 468 T-channels & (rules for \\
    & Learning + & & (channel + & action & top-K & no full & EM: 47\% (1-shot) & 446 A-channels & contrastive \\
    & LLM + & & functionality) & conditioned) & available) & retrain) & MRR@3: 0.65/0.45/0.50 & 1.3K T-func & learning + \\
    & Perplexity & & & & & & & 948 A-func & LLM LoRA) \\
    & Ranking & & & & & & & & \\
    \midrule
    % FARM (Ours)
    \rowcolor{gray!10}
    \textbf{FARM} &
    \textbf{Contrastive} &
    \textbf{NL query} &
    \textbf{Executable} &
    \textbf{\cmark} &
    \textbf{\cmark} &
    \textbf{\cmark} &
    \textbf{R@1: T72\%/A79\%, R@5: 92\%} &
    \textbf{16.5K applets} &
    \textbf{Medium} \\
    \rowcolor{gray!10}
    \textbf{(Ours)} &
    \textbf{Learning +} &
    &
    \textbf{JSON} &
    \textbf{(explicit)} &
    \textbf{(fallback)} &
    \textbf{(index)} &
    \textbf{MRR@5: 0.81, MRR@3: 0.84} &
    \textbf{1.7K trigger functions} &
    \textbf{(12.6K pairs} \\
    \rowcolor{gray!10}
    &
    \textbf{RAG + Multi-Agentic} &
    & & & & &
    \textbf{JM: 81\%/62\%/70\%} &
    \textbf{1.3K action functions} &
    \textbf{for encoder)} \\
    \rowcolor{gray!10}
    &
    \textbf{AI} &
    & & & & &
    \textbf{Faith: 0.44-0.48, Topic: 0.80-0.82} &
    & \\
    \bottomrule
    \end{tabular}
    }
    \vspace{2mm}

    \footnotesize
    \textbf{Legend:} \cmark = Has capability; \xmark = Does not have capability;
    N/A = Method does not produce executable outputs by design (recommendation systems).

    \textbf{Data Requirements:} FARM requires 16.5K applets (12.6K training pairs) for
    contrastive encoder fine-tuning and 3,011 Function-level schema descriptions for indexing (Applets).
    TARGE requires 34.8K rules for cross-view contrastive learning and LLM fine-tuning via LoRA.
    Other supervised methods require 45K-291K labeled (query $\rightarrow$ trigger + action) pairs
    or complex annotations.

    \textbf{FARM Metrics:} R@1 = Stage 1 retrieval (T=Trigger encoder, A=Action encoder reported separately); R@5 = Stage 1 retrieval (both encoders); MRR@5 = Stage 1 retrieval Mean Reciprocal Rank (macro-avg over trigger/action); MRR@3 = End-to-end Selection MRR; JM = Joint Accuracy on Gold/Noisy/One-shot sets; Faith = Faithfulness; Topic = Topic Adherence (Stage 2 quality).

    \textbf{Compositional Reasoning:} Explicit verification that trigger outputs match
    action inputs (FARM, TAPrec); Implicit through decoder architecture (RecipeGen);
    Partial through goal-based reasoning (TAPrec, TAP-AHGNN).

    \textbf{Service Extensibility:} Whether new services can be added without retraining
    the entire model. FARM supports this through simple index updates. 
    \normalsize
\end{table*}
\clearpage
%%%%%%%%%%%%%%%%%%%%%%%%%%%%%%%%%%%%%%%%%%%%%%%%%%%%%%%%%%%%%%%%%%%%%%%%%%%%%
\section{Approach}
\label{sec:approach}

In this section, we will introduce our purposed method, which comprises the following 2 stages which is illustrated in Figure~\ref{fig:system_architecture}:

%%%%%%%%%%%%%%%%%%%%%%%%%%%%%%%%%%%%%%%%%%%%%%%%%%%%%%%%%%%%%%%%%%%%%%%%%%%%%

%%%%%%%%%%%%%%%%%%%%%%%%%%%%%%%%%%%%%%%%%%%%%%%%%%%%%%%%%%%%%%%%%%%%%%%%%%%%%
\subsection{System Architecture Overview}
\label{subsec:overview}
Stage 1 uses contrastive-trained dual encoders for high-recall candidate retrieval. Stage 2 applies multi-agent selection with LLM-based verification to produce executable bindings. This leverages complementary strengths: neural speed and recall, plus LLM precision and schema-awareness.
\FloatBarrier   % <-- ADD THIS LINE

Formally, given a user query $q$, Stage 1 retrieves candidate sets:
\begin{align}
\mathcal{T}_k &= \underset{t_i \in \mathcal{T}}{\text{top-}k}\left(\text{sim}(\mathcal{E}_T(q), \mathcal{E}_T(t_i))\right) \\
\mathcal{A}_k &= \underset{a_j \in \mathcal{A}}{\text{top-}k}\left(\text{sim}(\mathcal{E}_A(q), \mathcal{E}_A(a_j))\right)
\end{align}

\noindent where $\mathcal{E}_T$ and $\mathcal{E}_A$ are the trigger and action encoders respectively, $\mathcal{T}$ and $\mathcal{A}$ are the full function catalogs, and $\text{sim}(\cdot, \cdot)$ denotes cosine similarity. Stage 2 then selects the optimal pair $(t^*, a^*) \in \mathcal{T}_k \times \mathcal{A}_k$ through multi-agent selection and generates the binding function $\beta: \mathcal{F}_a \rightarrow \mathcal{I}_t \cup \mathcal{S}$, mapping action input fields to trigger output ingredients or static values.

% ============================================================================
% FIGURE 1: High-Level System Architecture
% ============================================================================
\begin{figure*}[!b]
\centering
\includegraphics[width=1.0\linewidth]{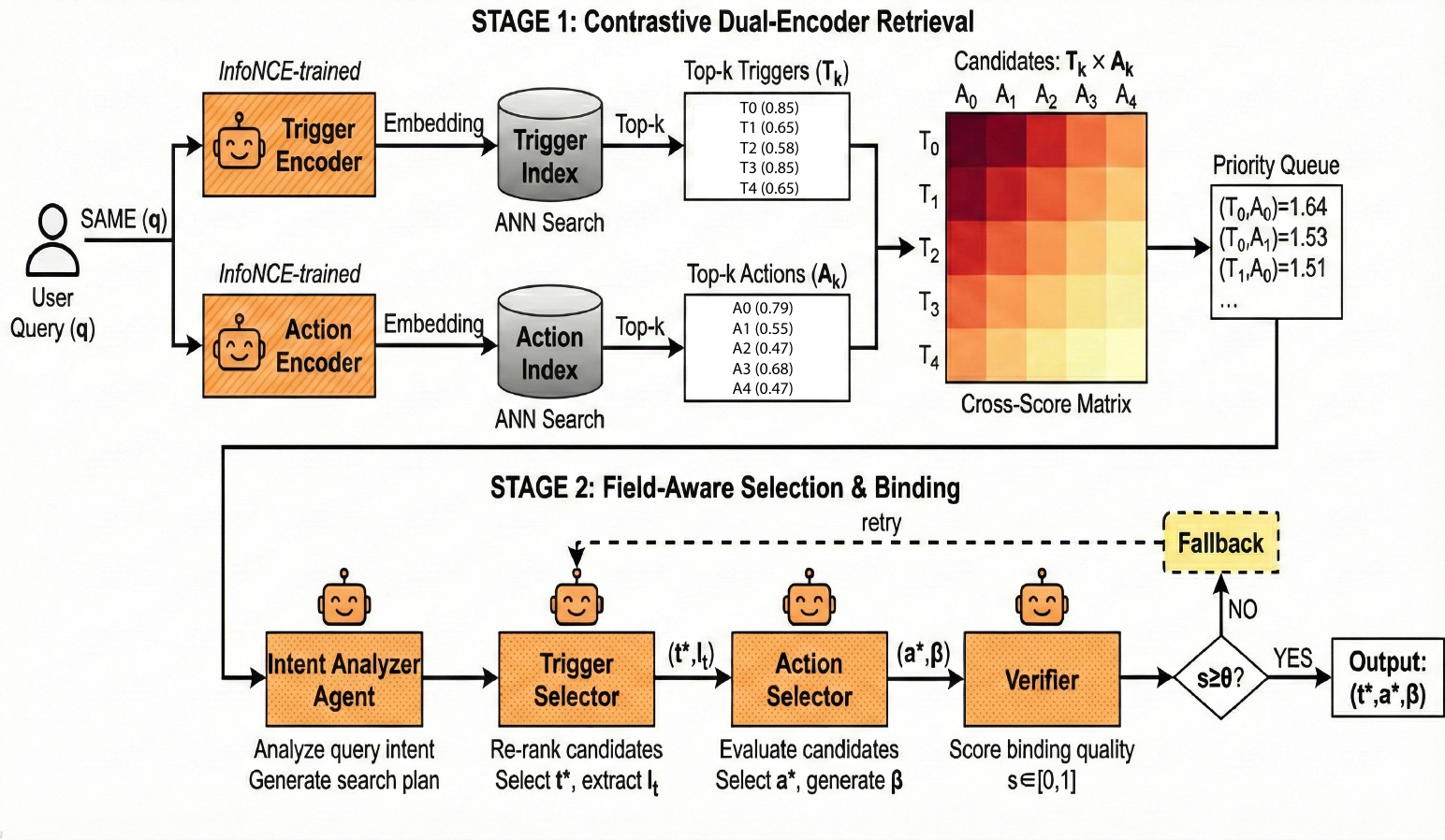}
\caption{FARM two-stage architecture. \textbf{Stage 1}: Dual contrastive
encoders retrieve top-$k$ trigger and action candidates; cross-score
matrix ranks $k \times k$ pairs by compatibility scores (70\% ingredient-field
coverage, 30\% retrieval quality). \textbf{Stage 2}:
Field-aware selection pipeline performs deep ingredient-to-field analysis—Intent Analyzer
decomposes query intent, Trigger Selector identifies optimal trigger $t^*$ and
extracts its ingredient schema $I_t$ (available output fields), Action Selector
evaluates field compatibility and generates bindings $\beta$ mapping ingredients
to action input fields, and Verifier scores binding completeness and semantic
coherence. Fallback retries with next candidate pair when verification score
$s < \theta$.}
\label{fig:system_architecture}\label{fig:system_architecture}
\end{figure*}

%%%%%%%%%%%%%%%%%%%%%%%%%%%%%%%%%%%%%%%%%%%%%%%%%%%%%%%%%%%%%%%%%%%%%%%%%%%%%
\subsection{Stage 1: Contrastive Dual-Encoder Retrieval}
\label{subsec:retrieval}

The retrieval stage addresses the scalability challenge by reducing the search space from $O(|\mathcal{T}| \times |\mathcal{A}|)$ to $O(k^2)$ where $k \ll |\mathcal{T}|, |\mathcal{A}|$. We train separate encoders for triggers and actions because the same query requires identifying different semantic aspects: the \emph{event} to detect (trigger) versus the \emph{action} to perform.

\subsubsection{Schema-Enriched Text Representation}

Unlike prior work that embeds only service names, we encode complete function schemas to enable field-level retrieval. Each function $x \in \mathcal{T} \cup \mathcal{A}$ is represented as:

\begin{equation}
\text{text}(x) = [\text{channel}]\ [\text{category}]\ \text{name}.\ \text{desc}\ ||\ \text{schema}(x)
\label{eq:text_repr}
\end{equation}

\noindent where $||$ denotes concatenation and $\text{schema}(x)$ encodes the function's data interface:

\textbf{Trigger Schema} (data provider):
\begin{equation}
\text{schema}(t) = \text{``Provides: ''} \bigoplus_{i \in \mathcal{I}_t} (\text{name}_i, \text{type}_i)
\end{equation}

\textbf{Action Schema} (data consumer):
\begin{equation}
\text{schema}(a) = \text{``Requires: ''} \bigoplus_{f \in \mathcal{F}_a} (\text{name}_f, \text{required}_f)
\end{equation}

\noindent where $\mathcal{I}_t$ denotes the ingredient set of trigger $t$, $\mathcal{F}_a$ denotes the field set of action $a$, and $\bigoplus$ represents formatted concatenation. This representation enables the model to learn that, for example, ``log sensor data'' should retrieve actions requiring data fields, not just services with ``log'' in the name.

\subsubsection*{Contrastive Learning with InfoNCE}

We train encoders using the InfoNCE objective~\cite{oord2018cpc}, which learns to discriminate positive query-function pairs from in-batch negatives:

\begin{equation}
\mathcal{L}_{\text{InfoNCE}} = -\mathbb{E}_{(q,d^+) \sim \mathcal{D}} \left[ \log \frac{\exp(s(q, d^+) / \tau)}{\sum_{i=1}^{B} \exp(s(q, d_i) / \tau)} \right]
\label{eq:infonce}
\end{equation}

\noindent where $s(q, d) = \cos(\mathcal{E}(q), \mathcal{E}(d))$ is the cosine similarity between query and document embeddings, $\tau$ is the temperature hyperparameter, $B$ is the batch size, and $d^+$ is the positive (ground-truth) function for query $q$. The denominator sums over all $B$ documents in the batch, treating co-occurring samples as hard negatives.

% ============================================================================
% FIGURE 2: InfoNCE Training and Dual Encoder Architecture
% ============================================================================
\begin{figure}[t]

\includegraphics[width=0.7\linewidth]{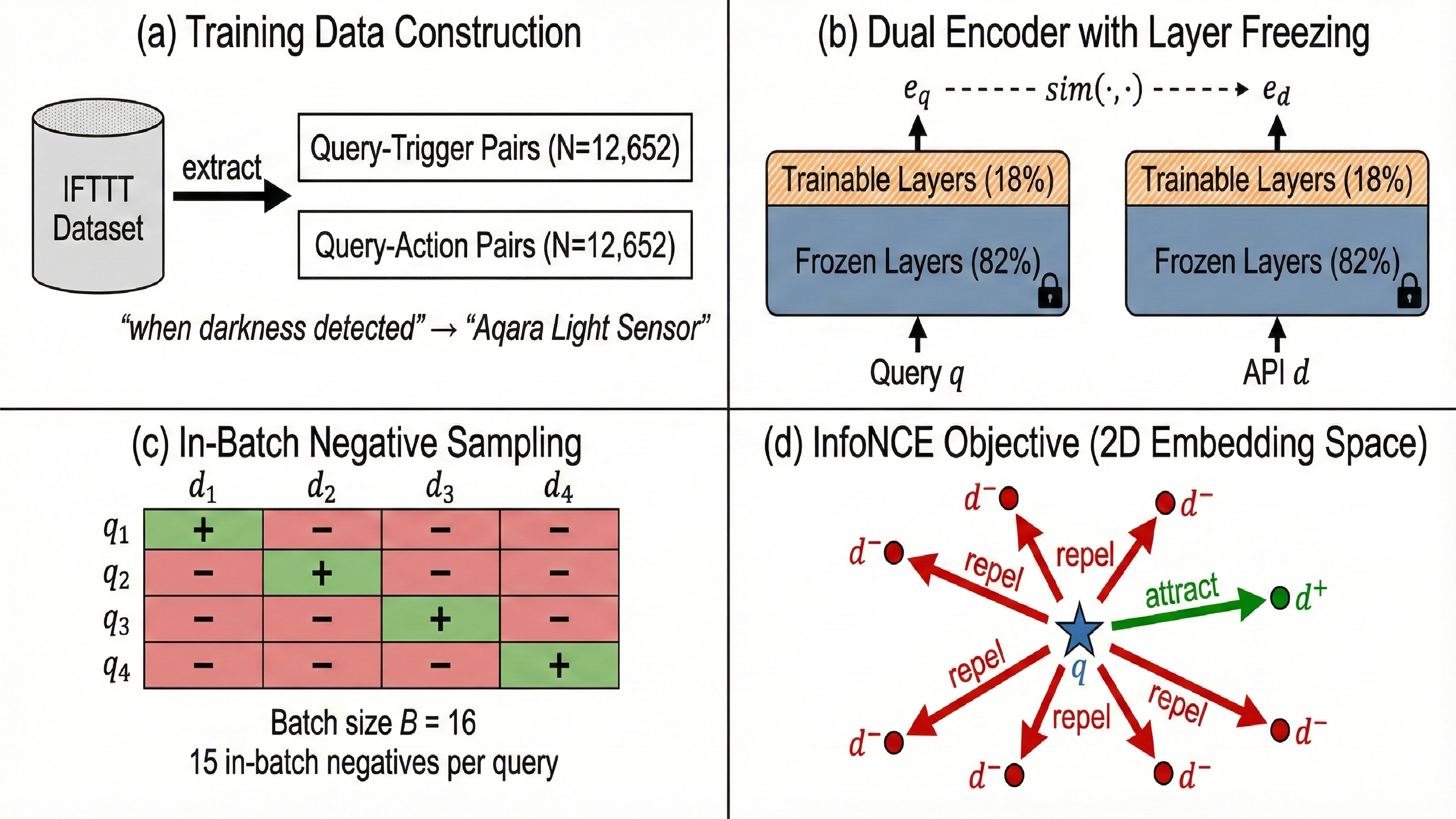}
\centering
% \fbox{\parbox{0.95\columnwidth}{\centering\vspace{5cm}

\caption{Contrastive training pipeline for dual encoders. \textbf{(a)} Training data construction: query-function pairs extracted from IFTTT dataset. \textbf{(b)} Dual encoder architecture with layer freezing—trainable layers (orange, 18\%) adapt to domain-specific patterns while frozen layers (blue-gray, 82\%) preserve pretrained semantics. \textbf{(c)} In-batch negative sampling: diagonal entries are positive pairs, off-diagonal entries serve as hard negatives. \textbf{(d)} InfoNCE objective in embedding space: queries are pulled toward positive documents and pushed away from negatives.}
\label{fig:infonce_training}
\end{figure}

% % ============================================================================
% % FIGURE 2: InfoNCE Training and Dual Encoder Architecture
% % ============================================================================
% \begin{figure}[t]
% \centering
% \fbox{\parbox{0.95\columnwidth}{\centering\vspace{5cm}
% \textbf{[FIGURE 2: Dual Encoder Training with InfoNCE Loss]}
% \vspace{5cm}}}

% \caption{Contrastive training of dual encoders. \textbf{(a)} Architecture: Query and API texts are encoded by shared-weight transformer with 82\% frozen layers. \textbf{(b)} InfoNCE objective: Positive pairs (green) are pulled together while in-batch negatives (red) are pushed apart in embedding space. \textbf{(c)} Layer freezing strategy: Lower layers (0-11) preserve pretrained semantics; upper layers (12-23) adapt to domain-specific retrieval.}
% \label{fig:infonce_training}
% \end{figure}

The temperature $\tau = 0.05$ creates a sharp softmax distribution that strongly penalizes confusion between similar candidates. With batch size $B = 16$, each training step provides 15 hard negatives sampled from the same domain, encouraging fine-grained discrimination.

\subsubsection{Semantic Preservation via Layer Freezing}
\label{subsubsec:layer_freezing}

A critical challenge in domain-specific training is catastrophic forgetting~\cite{kirkpatrick2017overcoming, li2016learning, zenke2017continual, delange2021continual}: the model loses pretrained semantic knowledge while adapting to the target task. We observed that full training caused the model to memorize exact service names while forgetting semantic equivalences (e.g., ''log'' $\approx$ ''record'' $\approx$ ``add row'').

We address this through selective layer freezing. Let $\theta = \{\theta^{(0)}, \theta^{(1)}, ..., \theta^{(L)}\}$ denote the parameters of an $L$-layer transformer. We partition layers into frozen and trainable sets:

\begin{equation}
\theta_{\text{frozen}} = \{\theta^{(0)}, ..., \theta^{(\ell)}\}, \quad \theta_{\text{train}} = \{\theta^{(\ell+1)}, ..., \theta^{(L)}\}
\end{equation}

\noindent During training, gradients are computed only for $\theta_{\text{train}}$:

\begin{equation}
\theta_{\text{train}}^{(t+1)} = \theta_{\text{train}}^{(t)} - \eta \nabla_{\theta_{\text{train}}} \mathcal{L}_{\text{InfoNCE}}
\end{equation}

For our final transformer encoder, we have chosen EmbeddingGemma \cite{embedding_gemma_2025} with 24 layers. We froze layers 0-11 (including the embedding layer), resulting in 82\% of the parameters (252M of 307M) frozen. This preserves the pretrained model's semantic generalization in lower layers while allowing upper layers to learn domain-specific retrieval patterns.

\textbf{Theoretical Motivation.} Research on transformer representations~\cite{jawahar2019bert, tenney2019bert, liu2019linguistic} shows that lower layers encode lexical and syntactic features while upper layers encode task-specific semantics. By freezing lower layers, we preserve the model's ability to recognize semantic equivalences (``darkness detected'' $\approx$ ``low light sensed'') while training upper layers for trigger-action discrimination.

\subsubsection{Training Configuration}

We train separate encoders for triggers and actions with identical architectures but independent weights. Table~\ref{tab:training_config} summarizes the key hyperparameters. We use a low temperature ($\tau = 0.05$) to create sharp softmax distributions that strongly penalize confusion between similar candidates. With batch size 16, each training step provides 15 hard negatives sampled from the same domain, encouraging fine-grained discrimination between semantically similar functions.

\begin{table}[h]
\centering
\small
\begin{tabular}{ll}
\toprule
\textbf{Parameter} & \textbf{Value} \\
\midrule
Base Model & Pretrained Text Encoder \\
Frozen Layers & 0--11 (+ embeddings) \\
Loss Function & InfoNCE \\
Temperature $\tau$ & 0.05 \\
Batch Size & 16 \\
Learning Rate & $2 \times 10^{-5}$ \\
Epochs & 3 \\
Training Pairs & 12,652 \\
\bottomrule
\end{tabular}
\caption{Encoder training configuration.}
\label{tab:training_config}
\end{table}

%%%%%%%%%%%%%%%%%%%%%%%%%%%%%%%%%%%%%%%%%%%%%%%%%%%%%%%%%%%%%%%%%%%%%%%%%%%%%

\subsubsection{Trigger-Action Programming Formulation}

Trigger Action Programming TAP enables end users to create automation rules in the form IF trigger THEN action, following the paradigm of Ur et al. \cite{ur2016tapwild}. Formally, given a user query $q$ expressed in natural language, the system must:

\begin{enumerate}
    \item \textbf{Select} a trigger $t^* \in \mathcal{T}$ from the trigger catalog
    \item \textbf{Select} an action $a^* \in \mathcal{A}$ from the action catalog
    \item \textbf{Generate} bindings $\beta: \mathcal{F}_a \rightarrow \mathcal{I}_t \cup \mathcal{S}$ mapping action input fields to trigger output ingredients or static values
\end{enumerate}

Each trigger $t$ exposes a set of \emph{ingredients} $\mathcal{I}_t = \{i_1, i_2, ...\}$---typed output fields produced when the trigger fires (e.g., \texttt{Subject}, \texttt{Body} for an SMS trigger). Each action $a$ requires a set of \emph{fields} $\mathcal{F}_a = \{f_1, f_2, ...\}$---typed input parameters needed for execution (e.g., \texttt{To}, \texttt{Subject} for an email action). The binding function $\beta$ creates the data flow that makes applets executable.

\subsubsection{Why Retrieval Alone is Insufficient}
\label{subsubsec:rag_limitation}

While retrieval-augmented generation (RAG) has achieved success in many NLP tasks~\cite{lewis2020retrieval, izacard2022atlas}, several factors make pure retrieval insufficient for TAP:

\textbf{Selection Ambiguity.} High recall ensures correct functions appear in the candidate set, but does not identify \emph{which} candidate is correct. With high retrieval recall (Section~\ref{subsec:stage1_results} shows R@5 $>$ 92\% for both encoders), selecting the correct pair from $k \times k$ combinations---25 candidate pairs when $k=5$ (Fig.~\ref{fig:stage2_mechanism})---requires reasoning beyond embedding similarity. Naive selection that simply takes the rank-1 result from each encoder evaluates only 1 pair and achieves 58\% joint accuracy, whereas multi-agent search over all 25 pairs achieves 81\% (Section~\ref{subsec:ablation}).

\textbf{Schema Compatibility.} Retrieval operates on text similarity, not schema compatibility. A trigger providing \texttt{[temperature, humidity]} may rank highly for ``weather logging'' but be incompatible with an action requiring \texttt{[image\_url, caption]}. Understanding data flow between functions requires schema-level reasoning.

\textbf{Binding Generation.} Even with correct trigger-action pairs, generating executable applets requires mapping trigger ingredients to action fields---determining that \texttt{StockName} should fill the \texttt{row\_content} field. This semantic alignment task is beyond retrieval's capability.

\textbf{Ambiguity Resolution.} User queries often admit multiple valid interpretations. ``Turn on lights when I leave'' could use location triggers (GPS-based) or calendar triggers (schedule-based); an LLM can reason about context and user intent to select appropriately.

These limitations motivate our two-stage architecture: retrieval for efficient candidate generation, followed by multi-agent selection for reasoning and binding.

\subsubsection{Design Consideration: Why Not LoRA?}
\label{subsubsec:why_not_lora}

Low-Rank Adaptation (LoRA)~\cite{hu2022lora} has become the dominant parameter-efficient fine-tuning method, achieving impressive results across many tasks by learning low-rank update matrices $\Delta W = BA$ where $B \in \mathbb{R}^{d \times r}$ and $A \in \mathbb{R}^{r \times d}$ with rank $r \ll d$. Extensions like QLoRA~\cite{dettmers2024qlora} further reduce memory requirements through quantization.

However, for contrastive retrieval training, we found layer freezing superior to LoRA for three reasons:

\textbf{Semantic Preservation.} LoRA modifies all layers simultaneously through low-rank updates, which can subtly alter the pretrained semantic space. Layer freezing ensures that frozen layers retain pretrained representations, which we validate empirically in Section~\ref{subsec:ablation}.

\textbf{Representation Stability.} Contrastive learning requires stable representations for effective in-batch negative sampling. LoRA's distributed updates across all layers can destabilize the embedding space during training, whereas layer freezing maintains a stable foundation in lower layers while adapting upper layers.

\textbf{Computational Simplicity.} Layer freezing requires no additional adapter modules, reducing implementation complexity and inference overhead. The frozen/trainable partition is straightforward: gradients simply are not computed for frozen parameters.

Recent work by Biderman et al.~\cite{biderman2024lora} has shown that LoRA can underperform full fine tuning on certain tasks, and our domain, contrastive retrieval with small datasets, appears to be one where the inductive bias of layer freezing that preserves lower layer semantics is particularly beneficial.
%%%%%%%%%%%%%%%%%%%%%%%%%%%%%%%%%%%%%%%%%%%%%%%%%%%%%%%%%%%%%%%%%%%%%%%%%%%%%

%%%%%%%%%%%%%%%%%%%%%%%%%%%%%%%%%%%%%%%%%%%%%%%%%%%%%%%%%%%%%%%%%%%%%%%%%%%%%
% ============================================================================
% FIGURE 3: Multi-Agent Selection Details
% ============================================================================
\begin{figure*}[t]
\centering
\includegraphics[width=0.95\textwidth]{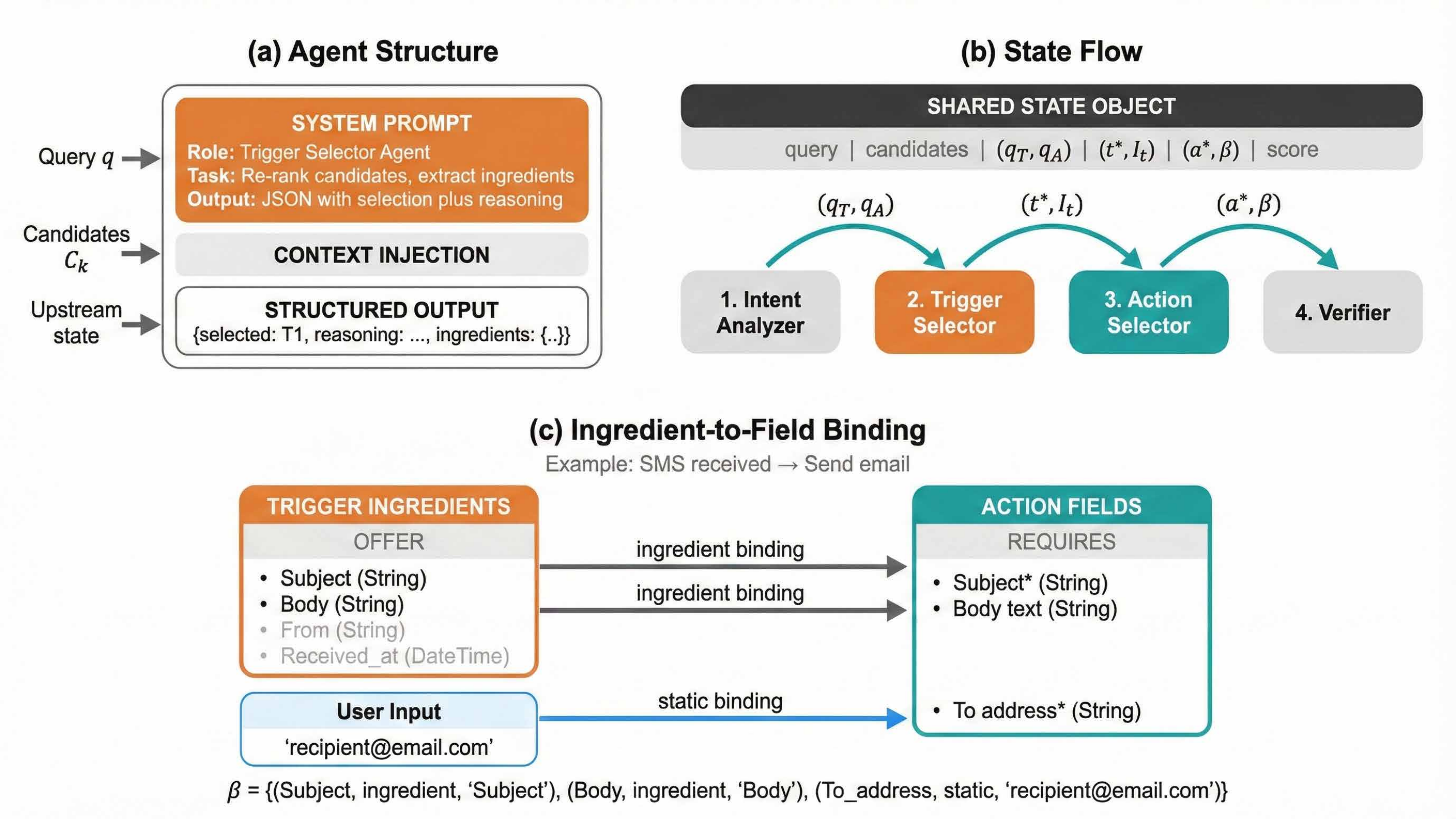}
\caption{Stage 2 multi-agent selection components. (a) Agent structure showing
input query, LLM processing with retrieved candidates $C_k$, and structured JSON
output. (b) State-based pipeline where Intent Analyzer, Trigger Selector, Action
Selector, and Verifier communicate through shared state. (c) Ingredient-to-field
binding illustrating how trigger outputs (Subject, Body) dynamically populate
action inputs, with static user-provided values for required fields.}
\end{figure*}

\subsubsection{State-Based Agent Communication}
\label{subsubsec:state_communication}

Unlike message-passing multi-agent systems where agents exchange discrete messages, our agents communicate through a \emph{shared state object} that accumulates information as execution proceeds. This design pattern, common in dataflow architectures, provides several advantages: (1) any downstream agent can access any upstream decision, (2) state can be serialized for debugging and replay, and (3) conditional routing decisions can inspect any state field.

Table~\ref{tab:state_schema} defines the key state fields and their producers/consumers.

\begin{table}[h]
\centering
\small
\begin{tabular}{lll}
\toprule
\textbf{State Field} & \textbf{Producer} & \textbf{Consumer} \\
\midrule
\texttt{query} $q$ & Input & All agents \\
\texttt{trigger\_candidates} $\mathcal{T}_k$ & Stage 1 & Trigger Selector \\
\texttt{action\_candidates} $\mathcal{A}_k$ & Stage 1 & Action Selector \\
\texttt{search\_intents} $(q_T, q_A)$ & Intent Analyzer & Selectors \\
\texttt{selected\_trigger} $(t^*, I_t)$ & Trigger Selector & Action Selector \\
\texttt{bindings} $(a^*, \beta)$ & Action Selector & Verifier \\
\texttt{verifier\_score} $s$ & Verifier & Routing logic \\
\texttt{llm\_overrode\_rag} & Selectors & Fallback logic \\
\bottomrule
\end{tabular}
\caption{State schema for inter-agent communication. Each agent reads upstream fields and writes to designated output fields.}
\label{tab:state_schema}
\end{table}

\subsubsection{Intent Analyzer Agent}
\label{subsubsec:analyzer}

The Analyzer Agent initiates Stage 2 by analyzing the user query using chain-of-thought (CoT) reasoning~\cite{wei2022chain}. Given query $q$, the agent performs:

\textbf{Intent Decomposition.} The analyzer identifies two distinct intents embedded in the query: (1) the trigger event to detect, and (2) the action to execute. For example, given ``Change the light to green if stock price rises,'' the analyzer extracts:
\begin{itemize}
    \item Trigger intent: ``stock price rises'' (financial market event)
    \item Action intent: ``change light to green'' (smart home control)
\end{itemize}

\textbf{Search Query Generation.} The analyzer generates optimized search queries $(q_T, q_A)$ to clarify trigger and action intents. These queries may expand abbreviations (``temp'' $\rightarrow$ ``temperature''), resolve ambiguities, or reformulate colloquial expressions into function-compatible terminology.

\textbf{Reasoning Trace.} Following the CoT paradigm~\cite{wei2022chainofthought, wang2024chainofthought}, the agent produces an explicit \texttt{THINKING} section followed by a \texttt{Intent Analyze}:
\begin{quote}
\small\ttfamily
THINKING: 1. User wants to detect stock price changes (TRIGGER).\\
2. User wants to control smart light color (ACTION).\\
3. Need financial function for trigger, IoT function for action.\\
PLAN: Search for stock/price triggers and light/color actions.
\end{quote}

\subsubsection{Trigger Selector Agent}

The Trigger Selector applies a two-stage selection process combining neural retrieval with LLM re-ranking, following the retrieve-then-rerank paradigm established in modern information retrieval~\cite{nogueira2019passage}.

\textbf{Candidate Pair Ranking.} Rather than selecting trigger and action independently, we pre-compute compatibility scores for all $k \times k$ candidate pairs:
\begin{equation}
\text{score}(t_i, a_j) = 0.7 \cdot \text{coverage}(t_i, a_j) + 0.3 \cdot \frac{\text{sim}(q, t_i) + \text{sim}(q, a_j)}{2}
\end{equation}
where $\text{coverage}(t_i, a_j)$ measures ingredient-to-field compatibility (defined below in Equation~\ref{eq:coverage}) and the retrieval similarity term provides ranking based on Stage 1 encoder quality. This weighted combination prioritizes schema compatibility (70\%) while using retrieval confidence (30\%) as a tie-breaker. Pairs are sorted by this score to create a priority queue enabling best-first search through the candidate space.

\textbf{Agreement-Based Selection.} The agent presents top-$k$ trigger candidates to the LLM with full function descriptions. The LLM selects its preferred match $t_{LLM}$ with reasoning. To prevent hallucination from degrading retrieval quality, we employ an \emph{agreement-based} mechanism:
\begin{equation}
t^* = \begin{cases}
t_{LLM} & \text{if } \dfrac{s_{RAG}(t_{LLM})}{s_{RAG}(t_0)} \geq \tau_T \\[8pt]
t_0 & \text{otherwise}
\end{cases}
\label{eq:agreement}
\end{equation}
where $t_0$ is RAG's top candidate, $s_{RAG}(\cdot)$ denotes retrieval similarity, and $\tau_T = 0.95$ is the trigger agreement threshold. This ensures LLM can only override RAG when its choice has comparable retrieval confidence, balancing reasoning capability with retrieval reliability.

\textbf{Ingredient Extraction.} After selection, the agent extracts available ingredients $I_t = \{(name, type, example)\}$ from the trigger's function schema. Ingredients represent data fields emitted when the trigger fires (e.g., \texttt{StockName}, \texttt{Price}, \texttt{PercentageChange}).

\subsubsection{Action Selector Agent}

The Action Selector receives the selected trigger $(t^*, I_t)$ and evaluates action candidates for schema compatibility.

\textbf{Cross-Scoring.} The agent computes a coverage score measuring how well trigger ingredients can satisfy action requirements:
\begin{equation}
\text{coverage}(t, a) = \frac{|\{f \in \mathcal{F}_a^{req} : \exists\, i \in I_t,\, \text{match}(i, f)\}|}{|\mathcal{F}_a^{req}|}
\label{eq:coverage}
\end{equation}
where $\mathcal{F}_a^{req}$ denotes required action fields. The $\text{match}(i, f)$ function checks compatibility using:
\begin{itemize}
    \item \textbf{Direct matching}: \texttt{temperature} $\rightarrow$ \texttt{temperature}
    \item \textbf{Substring matching}: \texttt{stock\_name} $\rightarrow$ \texttt{name}
    \item \textbf{Semantic mapping}: \texttt{message} $\approx$ \texttt{body} $\approx$ \texttt{content}
\end{itemize}

\textbf{LLM Re-ranking.} Similar to trigger selection, the LLM re-ranks action candidates with full context about the selected trigger and its ingredients. The agreement threshold $\tau_A = 0.80$ is lower than triggers, as action selection benefits more from LLM reasoning about functional intent.

\textbf{Binding Generation.} Upon selecting action $a^*$, the agent generates bindings $\beta$ mapping each action field to a data source:
\begin{equation}
\beta = \{(f, \text{src}, \text{val}) : f \in \mathcal{F}_a\}
\end{equation}
where $\text{src} \in \{\texttt{ingredient}, \texttt{static}\}$. Ingredient bindings use trigger data (e.g., \texttt{StockName} $\rightarrow$ \texttt{row\_content}); static bindings use placeholder values for fields without matching ingredients.

\subsubsection{Verifier Agent}

The Verifier implements the LLM-as-Judge pattern~\cite{zheng2023judging} to evaluate the complete applet configuration $(t^*, a^*, \beta)$:

\begin{itemize}
    \item \textbf{Binding Quality}: Are bindings semantically appropriate?
    \item \textbf{Completeness}: Are all required action fields bound?
    \item \textbf{Executability}: Can this configuration execute without runtime errors?
\end{itemize}

The Verifier produces a quality score $s \in [0, 1]$ and natural language critique. A rule-based fallback handles LLM parsing failures by checking trigger/action presence, binding count, and required field coverage.

\subsubsection{Quality-Gated Fallback}

When the verifier score falls below threshold $\theta_v = 0.5$, the system triggers fallback logic:

\begin{enumerate}
    \item \textbf{LLM Fallback}: If the LLM overrode RAG's selection (\texttt{llm\_overrode\_rag} = true), retry with RAG's original top choice. This catches cases where LLM reasoning was incorrect.
    \item \textbf{Pair Iteration}: If still failing, advance to the next candidate pair from the priority queue and restart from Trigger Selector.
\end{enumerate}

The system attempts up to $k^2$ pairs before declaring failure. In our evaluation (Section~\ref{subsec:stage2_results}), most queries succeed within two attempts due to Stage 1's high recall.

\subsubsection{Execution Example}
\label{subsubsec:execution_example}

We illustrate the complete pipeline with query: ``Change light to green if stock price rises.''

\textbf{Step 1: Analyzer.} Analyzes query and decomposes: $q_T$ = ``stock price rises,'' $q_A$ = ``change light color.'' Produces reasoning trace explaining the decomposition.

\textbf{Step 2: Trigger Selector.} Receives candidates $\mathcal{T}_5$ from RAG. Top results: $T_0$ = ``Price rises above'' (0.753), $T_1$ = ``Today's price rises by percentage'' (0.744). LLM selects $T_1$ with reasoning about percentage-based detection being more appropriate. Agreement ratio $0.744/0.753 = 0.99 \geq 0.95$, so LLM override accepted. Extracts $I_t$ = [\texttt{StockName}, \texttt{Price}, \texttt{PercentageChange}, \texttt{CheckTime}].

\textbf{Step 3: Action Selector.} Evaluates $\mathcal{A}_5$ against $I_t$. Computes coverage scores. Selects $A_0$ = ``Turn on / change light mode'' (0.627). Generates bindings: $\beta$ = [(\texttt{color}, static, ``green''), (\texttt{light}, static, ``Living room'')].

\textbf{Step 4: Verifier.} Evaluates $(t^*, a^*, \beta)$. Confirms all required fields bound. Assigns score $s = 0.85 \geq 0.5$. Output accepted.

\subsubsection{Chain-of-Thought Prompting}

All agents employ chain-of-thought (CoT) prompting~\cite{chiang2023chatbot} to produce interpretable reasoning traces. Each prompt instructs the LLM to:
\begin{enumerate}
    \item Analyze user intent explicitly before making decisions
    \item Evaluate each candidate's fitness with stated criteria
    \item Check schema compatibility between data sources and targets
    \item Justify the final selection with concrete reasoning
\end{enumerate}
This structured deliberation improves decision accuracy through explicit reasoning and provides interpretability for system debugging and error analysis.

%%%%%%%%%%%%%%%%%%%%%%%%%%%%%%%%%%%%%%%%%%%%%%%%%%%%%%%%%%%%%%%%%%%%%%%%%%%%%
\subsection{Embedding Model Selection}
\label{subsec:embedding_model}

The choice of base encoder significantly impacts retrieval quality. We identify three key selection criteria:

\textbf{Semantic Understanding.} The encoder should be trained on diverse corpora to provide robust semantic representations that generalize to WoT service descriptions. Models trained primarily on narrow domains may struggle with the mixed technical vocabulary common in function documentation.

\textbf{Efficiency.} The model should balance performance with computational tractability for domain-specific training. Larger models often offer diminishing returns, as pretrained general knowledge matters less than task-specific adaptation.

\textbf{Embedding Dimensionality.} The embedding dimension should balance expressiveness with retrieval efficiency, enabling fast approximate nearest neighbor search over large function catalogs.

For our implementation, we select Gemma Embedding~\cite{embedding_gemma_2025} (~307M parameters, 768-dimensional embeddings), which satisfies these criteria while remaining efficient for contrastive training.

\section{Experiments}
\label{sec:experiments}

We evaluate FARM through comprehensive experiments designed to assess both stages of our architecture. Our evaluation addresses three key questions: (1) Does contrastive training with layer freezing improve retrieval quality while preserving semantic generalization? (2) Does multi-agent selection improve end-to-end accuracy over retrieval alone? (3) How does FARM compare to existing trigger-action programming approaches?

%%%%%%%%%%%%%%%%%%%%%%%%%%%%%%%%%%%%%%%%%%%%%%%%%%%%%%%%%%%%%%%%%%%%%%%%%%%%%
\subsection{Function-Level TAP Dataset}
\label{subsec:dataset}

Our dataset comprises 1,724 trigger functions and 1,287 action functions from the IFTTT platform, spanning 19 categories including Smart Home, Finance, Health, Developer Tools, Social Media, and others. This creates a search space of $1,724 \times 1,287 = 2,218,788$ possible trigger-action pairs. Critically, our dataset operates at the \emph{function level} with complete schema specifications, which differs fundamentally from traditional service-level identification tasks.

\subsubsection{Service-Level vs. Function-Level Granularity}

To understand this distinction, consider the difference between identifying a service and selecting a specific trigger or action:

\begin{itemize}
    \item \textbf{Service-Level (Service):} Identify the service providing functionality (e.g., ``The New York Times'', ``Google Sheets'', ``Evernote'')
    \item \textbf{function-Level (Trigger/Action):} Select the specific trigger or action within that service (e.g., ``New article from search'', ``Add row to spreadsheet'', ``Append to note'')
\end{itemize}

A single service may expose dozens of distinct trigger and action functions. For instance, the New York Times service provides multiple trigger functions (``New article from search'', ``New popular article'', ``New article in section'', etc.), each with different functionality, parameters, and data outputs. Service-level identification only determines \emph{which service} to use. Functions-level selection determines \emph{which specific trigger or action} to configure and execute.

\subsubsection{Schema Structure and Data Interface Specifications}

Each trigger or action function in our dataset includes complete schema specifications that define its data interface. Table~\ref{tab:dataset_example} shows a concrete example from our dataset, illustrating the full schema structure for a trigger-action pair.:

\textbf{Trigger Schema:}
\begin{itemize}
    \item \texttt{service\_name}: Function identifier (e.g., ``New article from search'')
    \item \texttt{category}: Domain category (e.g., ``News \& information'')
    \item \texttt{description}: Natural language explanation of trigger behavior
    \item \texttt{Trigger fields}: Input parameters required to configure the trigger (e.g., ``Search for'' with type String)
    \item \texttt{Ingredients}: Output data fields produced when the trigger fires, with:
    \begin{itemize}
        \item \texttt{Slug}: Programmatic identifier (e.g., \texttt{Title}, \texttt{ArticleUrl})
        \item \texttt{Type}: Data type (String, Date, Number, Boolean, etc.)
        \item \texttt{Filter code}: Access path used by the IFTTT filter code environment
        \item \texttt{Example}: Representative value showing expected format
    \end{itemize}
\end{itemize}

\textbf{Action Schema:}
\begin{itemize}
    \item \texttt{service\_name}: Function identifier (e.g., ``Append to note'')
    \item \texttt{category}: Domain category (e.g., ``Popular services'')
    \item \texttt{description}: Natural language explanation of action behavior
    \item \texttt{Action fields}: Input parameters required to execute the action, with:
    \begin{itemize}
        \item \texttt{Label}: Human-readable field name (e.g., ``Title'', ``Body'')
        \item \texttt{Slug}: Programmatic identifier (e.g., \texttt{title}, \texttt{body})
        \item \texttt{Required}: Boolean indicating if field must be provided
        \item \texttt{Helper text}: Usage instructions or constraints
        \item \texttt{Filter code method}: Method signature used by the IFTTT filter code environment to set the field value
    \end{itemize}
\end{itemize}

\begin{table*}[t]
\centering
\small
\begin{tabular}{p{0.48\textwidth}p{0.48\textwidth}}
\toprule
\textbf{Trigger Function: ``New article from search''} & \textbf{Action Function: ``Append to note''} \\
\midrule
\textbf{Category:} News \& information & \textbf{Category:} Popular services \\
\midrule
\textbf{Description:} This Trigger fires every time a new article that is published by The New York Times matches a search query you specify. & \textbf{Description:} This Action will append to a note as determined by its title and notebook. \\
\midrule
\textbf{Trigger Fields:} & \textbf{Action Fields:} \\
$\bullet$ Search for (String, required) & $\bullet$ Title (String, required) \\
 & $\bullet$ Body (String, required) \\
 & $\bullet$ Notebook (String, optional) \\
 & $\bullet$ Tags (String, optional) \\
\midrule
\textbf{Ingredients (Outputs):} & \textbf{Data Flow Mapping:} \\
$\bullet$ Title (String) & Ingredients $\rightarrow$ Fields \\
$\bullet$ Author (String) & \\
$\bullet$ Blurb (String) & Example binding: \\
$\bullet$ ArticleUrl (String) & $\bullet$ Ingredient \texttt{Title} $\rightarrow$ Field \texttt{Title} \\
$\bullet$ ImageUrl (String) & $\bullet$ Ingredient \texttt{Blurb} $\rightarrow$ Field \texttt{Body} \\
$\bullet$ Source (String) & $\bullet$ Ingredient \texttt{Keywords} $\rightarrow$ Field \texttt{Tags} \\
$\bullet$ Section (String) & \\
$\bullet$ Keywords (String) & \\
$\bullet$ PublishedDate (Date with time) & \\
\bottomrule
\end{tabular}
\caption{Example function-Level Schema from Dataset. The trigger provides 9 typed ingredients (Title, Author, Blurb, ArticleUrl, ImageUrl, Source, Section, Keywords, PublishedDate); the action requires 4 fields (Title, Body, Notebook, Tags), with Title and Body marked as required. This schema information enables reasoning about data flow compatibility and automatic generation of ingredient-to-field bindings.}
\label{tab:dataset_example}
\end{table*}

\subsubsection{Why Function-Level Schemas Enable Executable Applets}

The distinction between service-level and function-level is not merely semantic. It fundamentally changes the task complexity and system requirements.

\textbf{Service-Level Limitation:} Prior work treats TAP as a classification problem over service names. A system that outputs ``The New York Times'' and ``Evernote'' provides no actionable information. Which specific trigger should be used. Which specific action should be selected. How to configure the parameters. Which ingredient values should fill which action fields. The output requires extensive manual configuration before execution.

\textbf{Function-Level Advantage:} Our dataset's function-level schemas enable fully automated applet generation. By including:
\begin{itemize}
    \item \textbf{Function specifications}: The exact trigger and action functions to select (``New article from search'', ``Append to note'')
    \item \textbf{Type information}: Data type constraints for validation (String, Date, Boolean)
    \item \textbf{Required field markers}: Which parameters must be provided versus optional
    \item \textbf{Ingredient definitions}: Available data outputs from triggers
    \item \textbf{Field requirements}: Expected inputs for actions
\end{itemize}

The system can reason about \emph{schema compatibility} (Does the trigger produce data the action needs?), generate \emph{ingredient-to-field bindings} (Which trigger outputs map to which action inputs?), and produce \emph{executable configurations} ready for deployment without human intervention.

For example, given the query ``Send New York Times articles about recipes to Evernote'', a service-level system outputs service names, while our function-level system produces:
\begin{itemize}
    \item Trigger: ``New article from search'' with field \texttt{Search for} = ``recipe''
    \item Action: ``Append to note'' with bindings:
    \begin{itemize}
        \item \texttt{Title} $\leftarrow$ Ingredient \texttt{Title}
        \item \texttt{Body} $\leftarrow$ Ingredient \texttt{Blurb}
        \item \texttt{Tags} $\leftarrow$ Ingredient \texttt{Keywords}
    \end{itemize}
\end{itemize}

This complete specification is executable within the IFTTT execution model without further configuration. The function-level granularity with schema information is what makes our task and dataset fundamentally different from prior service identification approaches.

We evaluate FARM on the IFTTT platform, a widely-used trigger-action programming service for WoT (Web of Things) automation. The service catalog comprises:

\begin{itemize}
    \item \textbf{Triggers}: 1,724 distinct functions across 19 categories (Smart Home, Finance, Health, Developer Tools, Social Media, etc.)
    \item \textbf{Actions}: 1,287 distinct functions across 19 categories
    \item \textbf{Search Space}: $1,724 \times 1,287 = 2,218,788$ possible trigger-action function pairs
\end{itemize}

Each function entry contains structured schema information: function name, category, natural language description, and data interface specifications (ingredients for triggers, required fields for actions). Following IFTTT's internal terminology, the dataset labels specific functions as \texttt{service\_name} (e.g., ``New article from search'', ``Add row to spreadsheet''), which we evaluate at the function-level rather than the platform-level (e.g., ``The New York Times'', ``Google Sheets'').

\paragraph{Terminology.} IFTTT refers to individual trigger/action functions as \texttt{service\_name} (e.g., ``Add row to spreadsheet''). We refer to these as \emph{functions} throughout. We reserve \emph{platform service} for the parent application (e.g., ``Google Sheets''). Unless stated otherwise (e.g., Table~\ref{tab:baseline_comparison}), all Stage 1 and Stage 2 metrics are computed at the function level.

%%%%%%%%%%%%%%%%%%%%%%%%%%%%%%%%%%%%%%%%%%%%%%%%%%%%%%%%%%%%%%%%%%%%%%%%%%%%%
\subsection{Experimental Setup}
\label{subsec:setup}

\subsubsection{Evaluation Sets}

Following the evaluation protocol established by Cimino et al.~\cite{cimino2025targe}, we construct three evaluation sets to test different aspects of system robustness:

\begin{itemize}
    \item \textbf{Gold Set}: Clear, well-formed automation requests with unambiguous intent (e.g., ``When darkness detected, log to spreadsheet'')
    \item \textbf{Noisy Set}: Vague or ambiguous descriptions requiring inference (e.g., ``track my fitness stuff'')
    \item \textbf{One-Shot Set}: Queries involving rare or specialized functions with limited training exposure
\end{itemize}

\subsubsection{Model Configuration}

\textbf{Stage 1 - Contrastive Encoders:}
\begin{itemize}
    \item \textbf{Base Model}: EmbeddingGemma~\cite{embedding_gemma_2025} (~307M parameters, 768-dimensional embeddings)
    \item \textbf{Architecture}: 24 transformer layers, 12 attention heads, 2,048 token context
    \item \textbf{Layer Freezing}: Layers 0--11 frozen, layers 12--23 trainable (18\% of total parameters trainable)
    \item \textbf{Training}: InfoNCE loss~\cite{oord2018cpc}, $\tau=0.05$, batch size 16, 3 epochs, learning rate $2 \times 10^{-5}$
\end{itemize}

\textbf{Stage 2 - Multi-Agent Selection:}
\begin{itemize}
    \item \textbf{LLM}: IBM Granite 4.0 Small~\cite{ibm_granite4_announcement_2025} via local inference
    \item \textbf{Temperature}: 0.0 (deterministic outputs)
\item \textbf{Note}: The multi-agent system operates through \emph{prompting only}---no additional training is performed on the LLM. As Brown et al.\ \cite{brown2020language} demonstrated, large-scale language models can achieve strong task performance without fine-tuning, and Wei et al.\ \cite{wei2022emergent} showed that emergent abilities arise at sufficient scale through prompting alone. Agents use carefully designed system prompts with chain-of-thought reasoning~\cite{wei2022chainofthought} to elicit step-by-step analysis before producing structured outputs.\end{itemize}

%%%%%%%%%%%%%%%%%%%%%%%%%%%%%%%%%%%%%%%%%%%%%%%%%%%%%%%%%%%%%%%%%%%%%%%%%%%%%
\subsection{Evaluation Metrics}
\label{subsec:metrics}

We employ metrics from established benchmarks to enable fair comparison with prior work.

\subsubsection{Stage 1: Retrieval Metrics}

We evaluate retrieval effectiveness using Recall at cutoff $K$ (Recall@K) and Mean Reciprocal Rank at cutoff $K$ (MRR@K), two standard ranking metrics in information retrieval~\cite{voorhees1999trec,manning2008ir}.

\textbf{Recall@K} measures retrieval coverage: what fraction of queries have the correct answer in the top-$K$ results? Given a query set $Q$:
\begin{equation}
\mathrm{Recall@K} = \frac{1}{|Q|} \sum_{q \in Q} \mathbb{I}\left( \exists \, d \in \mathcal{D}_q^K \;\; \text{s.t.} \;\; d \in \mathcal{R}_q \right),
\end{equation}
where $\mathcal{D}_q^K$ denotes the top $K$ retrieved documents for query $q$, $\mathcal{R}_q$ is the set of relevant documents for $q$, and $\mathbb{I}(\cdot)$ is the indicator function. Practically, R@1 asks: ``Is the correct answer ranked first?'' while R@5 asks: ``Is the correct answer somewhere in the top 5?'' We report R@1 and R@5 to evaluate both precision (top result quality) and recall (candidate set coverage for Stage 2).

\textbf{Joint Recall@K} extends this to dual-encoder retrieval, measuring when \emph{both} components succeed simultaneously:
\begin{equation}
\mathrm{Joint\ Recall@K} = \frac{1}{|Q|} \sum_{q \in Q} \mathbb{I}\left( t_{\text{true}} \in \mathcal{D}_q^K(T) \land a_{\text{true}} \in \mathcal{D}_q^K(A) \right),
\end{equation}
where $\mathcal{D}_q^K(T)$ and $\mathcal{D}_q^K(A)$ denote the top $K$ triggers and actions retrieved for query $q$. For example, Joint R@5 = 85\% means that for 85\% of queries, the correct trigger appears in the trigger encoder's top-5 \emph{and} the correct action appears in the action encoder's top-5. Critically, Joint R@5 establishes the \emph{theoretical ceiling} for Stage 2 performance: Stage 2 selection cannot choose a correct pair that Stage 1 did not retrieve (Fig.~\ref{fig:stage2_mechanism}).

\textbf{Mean Reciprocal Rank (MRR@K)} measures ranking quality by considering \emph{where} the correct answer appears within the top-$K$ results:
\begin{equation}
\mathrm{MRR@K} = \frac{1}{|Q|} \sum_{q \in Q}
\begin{cases}
\frac{1}{\operatorname{rank}_q} & \text{if } \operatorname{rank}_q \le K,\\
0 & \text{if } \operatorname{rank}_q > K,
\end{cases}
\end{equation}
where $\operatorname{rank}_q$ is the rank position of the first relevant document for query $q$. MRR@K gives partial credit based on position: rank 1 receives score 1.0, rank 2 receives 0.5, rank 5 receives 0.2, while anything beyond rank $K$ receives 0. The cutoff $K$ defines the evaluation scope and we use two values: (1) \textbf{MRR@5} for Stage 1 retrieval evaluation, measuring how well individual encoders rank trigger and action candidates (macro-averaged over both encoders, reported in Fig.~\ref{fig:retrieval_quality}); (2) \textbf{MRR@3} for function-level prediction, applied \emph{twice}---first to evaluate FARM's own trigger-action pair selection performance, then to compare FARM against prior work (LAM~\cite{liu2016latent}, RecipeGen++~\cite{yusuf2022recipegentpp}, TARGE~\cite{cimino2025targe}) using the same metric for fair comparison (Table~\ref{tab:baseline_comparison}). MRR@3 follows established evaluation protocols in the TAP literature for function-level prediction tasks.

\subsubsection{Stage 2: Selection and Generation Metrics}

We define end-to-end evaluation metrics that measure system performance on complete applet generation:

\textbf{Goal Accuracy} measures task completion with partial credit, evaluating whether the system achieves the user's automation intent:
\begin{equation}
\text{Goal Accuracy}(q) = \begin{cases}
1.0 & \text{if } t_{\text{pred}} = t_{\text{true}} \land a_{\text{pred}} = a_{\text{true}} \\
0.5 & \text{if } (t_{\text{pred}} = t_{\text{true}}) \oplus (a_{\text{pred}} = a_{\text{true}}) \\
0.0 & \text{otherwise}
\end{cases}
\end{equation}
where $\oplus$ denotes exclusive OR (exactly one match). Practically, Goal Accuracy = 1.0 means the applet is fully correct, 0.5 means partial success (user achieves either trigger monitoring or action execution, but not the complete automation), and 0.0 means complete failure. We consider a query successful when Goal Accuracy $\geq 0.5$, used to compute the Success Rate metric.

\textbf{Joint Accuracy} measures service-pair correctness, requiring both trigger and action to be correctly selected:
$$
\text{Joint Accuracy}(q) = \mathbb{I}(t_{\text{pred}} = t_{\text{true}} \land a_{\text{pred}} = a_{\text{true}}),
$$
where $t_{\text{pred}}$ and $a_{\text{pred}}$ are the predicted trigger and action service names, $t_{\text{true}}$ and $a_{\text{true}}$ are the ground truth services, and $\mathbb{I}(\cdot)$ is the indicator function. Joint Accuracy returns 1 only when both services are correct, and 0 otherwise. It is computed from the \emph{full system} (Stage 1 + Stage 2 multi-agent selection), allowing Stage 2 LLM reasoning to refine Stage 1 candidates. Unlike Goal Accuracy which provides partial credit (0.5 for one service correct), Joint Accuracy is binary (0/1), making it stricter and comparable to metrics used in prior work (LAM, RecipeGen++, TARGE).

\textbf{Trigger/Action Accuracy} measures individual component correctness:
\begin{equation}
\text{Trigger Accuracy} = \frac{1}{|Q|} \sum_{q \in Q} \mathbb{I}(t_{\text{pred}} = t_{\text{true}}),
\end{equation}
and similarly for Action Accuracy with $a_{\text{pred}}$ and $a_{\text{true}}$. These metrics isolate which component (trigger selection vs. action selection) causes failures, informing system debugging and improvement.

\textbf{Success Rate} measures task-level success with partial credit, defined as the fraction of queries where the system achieves at least partial automation:
\begin{equation}
\text{Success Rate} = \frac{1}{|Q|}\sum_{q \in Q} \mathbb{I}(\text{Goal Accuracy}(q) \geq 0.5).
\end{equation}
Since $\text{Goal Accuracy}(q) \in \{0.0, 0.5, 1.0\}$ and $\text{Joint Accuracy}(q) \in \{0, 1\}$, the following ordering holds pointwise and therefore also in expectation:
\begin{equation}
\text{Success Rate} \geq \mathbb{E}[\text{Goal Accuracy}] \geq \mathbb{E}[\text{Joint Accuracy}].
\end{equation}
Practically, Success Rate counts queries where at least one service (trigger or action) is correctly identified. Success Rate $<$ 100\% indicates failures in service identification or system crashes.

We adopt generation quality metrics from the RAGAS evaluation framework~\cite{es2024ragas}, following the functional definitions provided in the official documentation~\cite{ragas_docs}.

\textbf{Faithfulness} measures whether generated applet configurations are grounded in retrieved function schemas, preventing hallucination:
\begin{equation}
\text{Faithfulness} = \frac{|\text{Claims verifiable in retrieved schemas}|}{|\text{Total claims in generated applet}|}.
\label{eq:faithfulness}
\end{equation}
A ``claim'' is an assertion about service capabilities, field names, or data types in the generated applet. For example, if the system generates a binding \texttt{trigger.ingredient\_name $\to$ action.field\_name}, Faithfulness checks whether both \texttt{ingredient\_name} exists in the trigger schema and \texttt{field\_name} exists in the action schema. Faithfulness = 1.0 means all generated content is verifiable from retrieved documentation; lower scores indicate hallucinated field names or capabilities not present in the actual function specifications.

\textbf{Topic Adherence} evaluates whether selected services match the user's intended automation domain:
\begin{equation}
\text{Topic Adherence} = \frac{|\text{Applets matching reference automation domain}|}{|\text{Total applets}|}.
\label{eq:topic_adherence}
\end{equation}
Practically, if a user's query mentions ``fitness tracking,'' Topic Adherence checks whether the selected trigger and action belong to health/fitness services (e.g., Fitbit, Google Fit, MyFitnessPal) rather than unrelated domains (e.g., social media, smart home). This metric ensures the system respects user intent boundaries and does not inappropriately cross automation contexts.

%%%%%%%%%%%%%%%%%%%%%%%%%%%%%%%%%%%%%%%%%%%%%%%%%%%%%%%%%%%%%%%%%%%%%%%%%%%%%
\subsection{Experiment results}
\subsubsection{Stage 1: Contrastive Learning Results}
\label{subsec:stage1_results}

\begin{figure}[htbp]
\centering
\includegraphics[width=0.80\columnwidth]{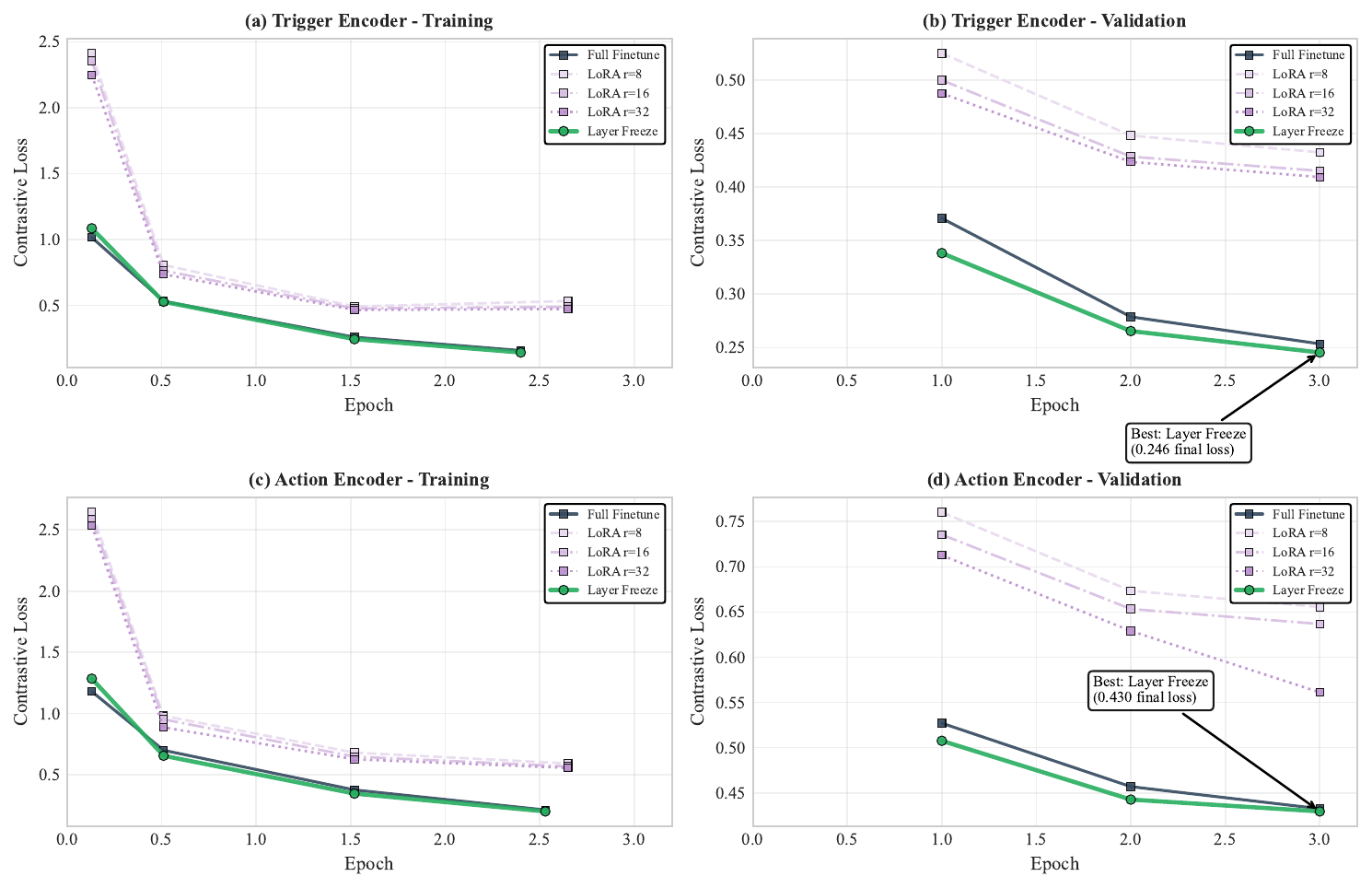}
\caption{Contrastive training curves showing loss convergence and R@1 improvement. Both encoders converge within 3 epochs, with the majority of improvement in the first epoch.}
\label{fig:training_curves}
\end{figure}

\begin{figure}[htbp]
\centering
\includegraphics[width=0.80\columnwidth]{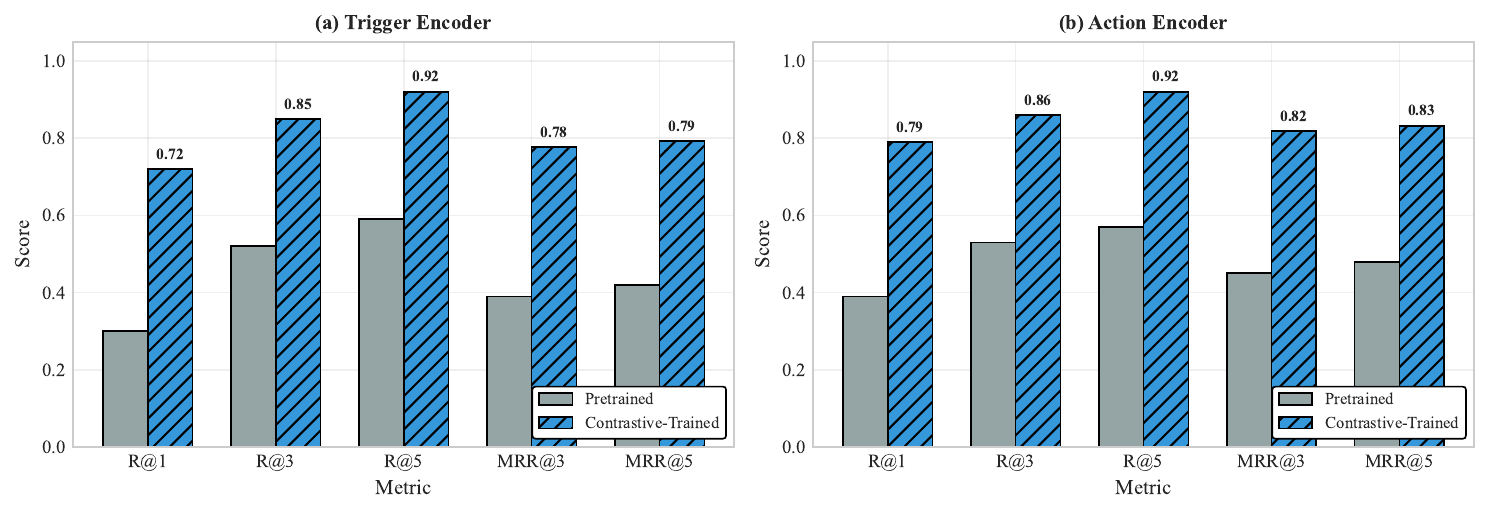}
\caption{Stage 1 retrieval quality. Contrastive training with layer freezing significantly improves trigger encoder (R@1: 30\%$\rightarrow$72\%, R@5: 59\%$\rightarrow$92\%, MRR@5: 42\%$\rightarrow$79\%) and action encoder (R@1: 39\%$\rightarrow$79\%, R@5: 57\%$\rightarrow$92\%, MRR@5: 48\%$\rightarrow$83\%).}
\label{fig:retrieval_quality}
\end{figure}

We first evaluate the retrieval stage in isolation to understand the impact of contrastive training and layer freezing. Figure~\ref{fig:training_curves} illustrates the training progression over three epochs. Figure~\ref{fig:retrieval_quality} presents retrieval performance comparing pretrained baseline against contrastive-trained encoders on the Gold evaluation set.

% \subsubsection{Training Dynamics}

Layer freezing with contrastive training achieves substantial improvements across all retrieval metrics. Trigger encoder: R@1 improves from 30\% to 72\% (+140\%), R@5 from 59\% to 92\% (+56\%), and MRR@5 from 42\% to 79\% (+88\%). Action encoder: R@1 improves from 39\% to 79\% (+103\%), R@5 from 57\% to 92\% (+61\%), and MRR@5 from 48\% to 83\% (+73\%). Both encoders achieve $>$92\% R@5, ensuring the target trigger/action almost always appears in the top-5 candidates for Stage 2 selection.

Critically, these individual R@5 scores correspond to a measured Joint R@5 of 85\%, meaning both the correct trigger \emph{and} correct action appear in their respective top-5 results for 85\% of queries. This Joint R@5 establishes the theoretical upper bound for Stage 2 performance---our multi-agent selection can only choose from pairs that Stage 1 successfully retrieved.
\FloatBarrier

%%%%%%%%%%%%%%%%%%%%%%%%%%%%%%%%%%%%%%%%%%%%%%%%%%%%%%%%%%%%%%%%%%%%%%%%%%%%%
\subsubsection{Stage 2: Multi-Agent Selection Results}
\label{subsec:stage2_results}

\begin{figure}[!htbp]
\centering
\includegraphics[width=0.80\columnwidth]{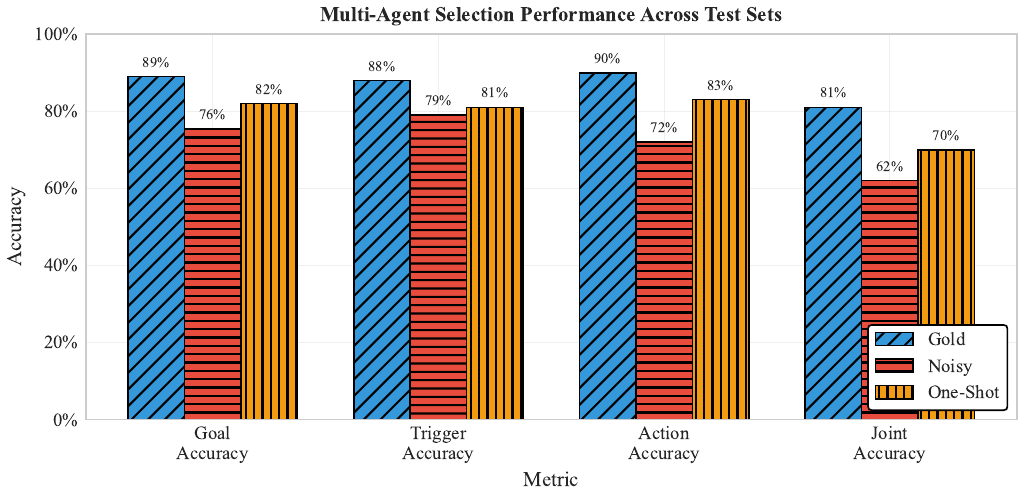}
\caption{Stage 2 multi-agent selection performance across evaluation sets. The system achieves 81\% joint accuracy on Gold, with graceful degradation on challenging inputs.}
\label{fig:selection_performance}
\end{figure}

We evaluate the complete two-stage system, measuring end-to-end selection accuracy and generation quality across all three evaluation sets (Gold, Noisy, and One-Shot).
Figure~\ref{fig:selection_performance} presents selection accuracy across all three evaluation sets.

\textbf{Gold Set:} Clear queries enable high-confidence selection. The gap between individual accuracy (89-90\%) and joint accuracy (81\%) reflects the compounding difficulty of selecting both services correctly.

\textbf{Noisy Set:} Ambiguous queries challenge the system, particularly for action selection (72\%). Vague descriptions require significant inference.

\textbf{One-Shot Set:} The system maintains reasonable performance on rare functions, demonstrating robust generalization from contrastive learning.

\subsubsection{Generation Quality}

\begin{figure}[!htbp]
\centering
\includegraphics[width=0.80\columnwidth]{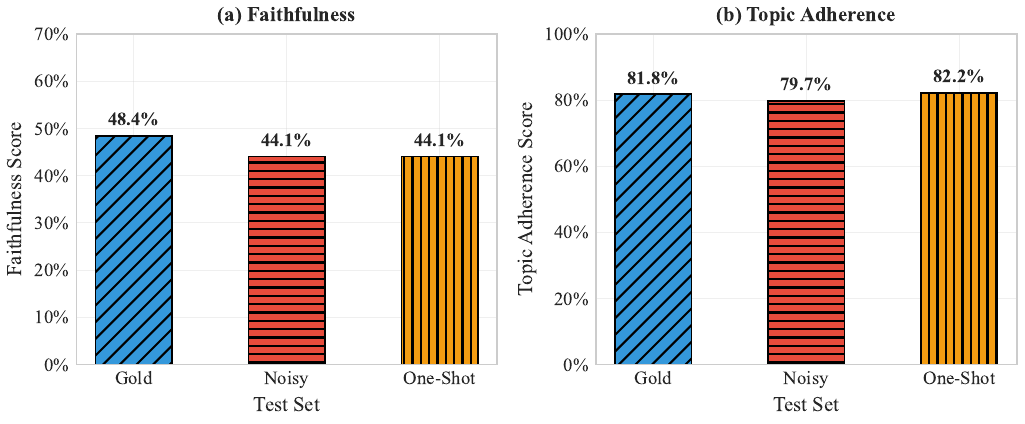}
\caption{Generation quality metrics across evaluation sets. Faithfulness (0.44--0.48) measures grounding in retrieved schemas; Topic Adherence (0.80--0.82) measures alignment with user's automation domain.}
\label{fig:quality_metrics}
\end{figure}

Figure~\ref{fig:quality_metrics} presents RAGAS~\cite{es2024ragas} quality metrics for generated applet configurations.

High topic adherence across all sets indicates that selected services consistently align with the user's stated automation domain. Moderate faithfulness scores reflect the challenging nature of generating precise field bindings that require inference beyond explicit context.

\subsubsection{Comprehensive Multi-Dataset Evaluation}

\begin{figure}[!htbp]
\centering
\includegraphics[width=0.95\columnwidth]{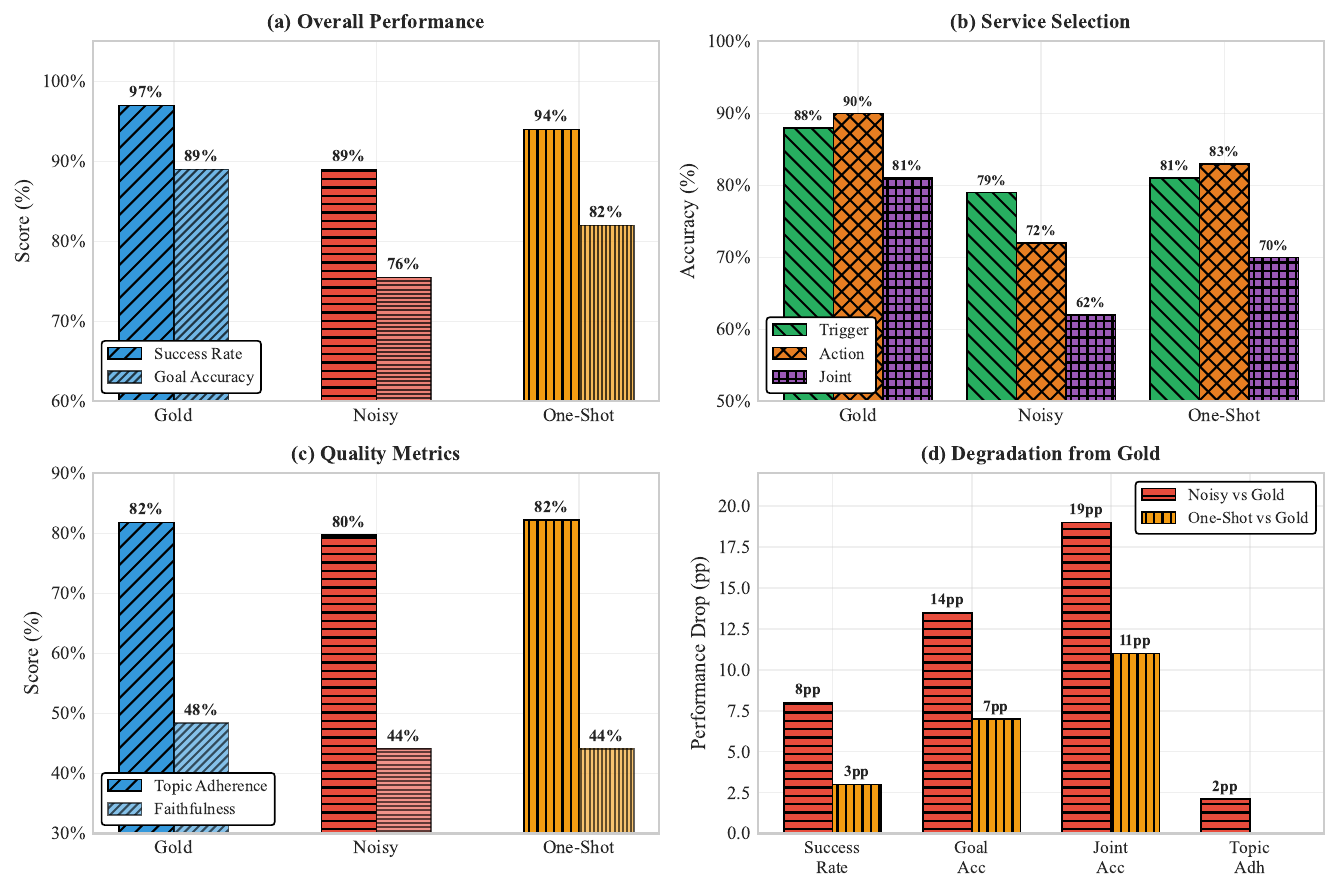}
\caption{Multi-agentic evaluation across test sets. \textbf{(a)} Overall performance: Gold achieves 97\% success rate (at least one service correct) and 89\% goal accuracy (average with partial credit); Noisy shows 8pp degradation in success rate; One-Shot maintains 94\% success. The ordering Success Rate $>$ Goal Accuracy reflects metric definitions: Success Rate = binary threshold (goal\_accuracy $\geq$ 0.5), Goal Accuracy = arithmetic mean. \textbf{(b)} Service selection: Trigger accuracy (88\%/79\%/81\%), Action accuracy (90\%/72\%/83\%), Joint accuracy (81\%/62\%/70\%) for Gold/Noisy/One-Shot respectively. \textbf{(c)} Quality metrics: Topic adherence remains stable (82\%) across sets; Faithfulness moderate (44--48\%). \textbf{(d)} Performance degradation: Noisy queries show largest drops in joint accuracy (19pp) while One-Shot maintains resilience (11pp drop).}
\label{fig:comprehensive_evaluation}
\end{figure}

Figure~\ref{fig:comprehensive_evaluation} presents a comprehensive breakdown of system performance across all three evaluation sets, revealing both overall effectiveness and robustness characteristics.

\textbf{Gold Set Results:} Clear, well-formed queries enable high performance across all metrics. The system achieves 97\% success rate (at least partial automation achieved), 89\% goal accuracy (average function-level performance with partial credit), and 81\% joint accuracy (both trigger and action functions correctly selected). Individual component accuracy reaches 88\% for triggers and 90\% for actions, with the gap to joint accuracy (81\%) reflecting the compounding difficulty of perfect service-pair selection.

\textbf{Noisy Set Results:} Vague or ambiguous descriptions present significant challenges. Success rate drops to 89\% (8pp degradation from Gold), goal accuracy to 75.5\% (13.5pp drop), and joint accuracy to 62\% (19pp drop). The ordering Success Rate (89\%) $>$ Goal Accuracy (75.5\%) $>$ Joint Accuracy (62\%) reflects the metric hierarchy: Success Rate requires only partial correctness (one service), Goal Accuracy averages partial credit, while Joint Accuracy requires both services correct. Action selection proves particularly challenging (72\% accuracy vs 90\% on Gold), as ambiguous queries require substantial inference to identify appropriate actions. Topic adherence remains stable at 79.7\%, indicating robust domain alignment despite query ambiguity.

\textbf{One-Shot Set Results:} Queries involving rare functions demonstrate robust generalization from contrastive learning. The system maintains 94\% success rate (3pp drop from Gold) and 82\% goal accuracy (7pp drop), with 70\% joint accuracy (11pp drop). Both trigger (81\%) and action (83\%) selection remain competitive, showing that layer freezing preserves sufficient pretrained semantic knowledge to handle low-resource scenarios.

\textbf{Quality Metrics Analysis:} Faithfulness scores range from 44\% to 48\% across all sets, reflecting the challenging nature of schema-grounded binding generation that requires inference beyond explicit context. Topic adherence remains consistently high (80--82\%), indicating that the multi-agent system reliably selects services within the user's intended automation domain across all test conditions.

%%%%%%%%%%%%%%%%%%%%%%%%%%%%%%%%%%%%%%%%%%%%%%%%%%%%%%%%%%%%%%%%%%%%%%%%%%%%%
\subsection{Comparison with Related Work}
\label{subsec:comparison}

We compare FARM against three established baselines for trigger-action programming:
Liu et al.~\cite{liu2016latent}, Yusuf et al.~\cite{yusuf2022recipegentpp}, and
Cimino et al.~\cite{cimino2025targe}. Table~\ref{tab:baseline_comparison} presents
quantitative results on a service-level prediction task across all three evaluation
sets. Although FARM operates at function-level granularity with full schema
specifications, for this table we map each predicted function to its parent
platform service to enable fair comparison with baselines. To ensure a controlled
comparison, we train and evaluate each baseline on the same service-level splits
derived from our dataset.

\begin{table*}[h]
\centering
\vspace{2mm}  % Add padding from top
\begin{tabular}{l|cc|cc|cc}
\toprule
& \multicolumn{2}{c|}{\textbf{Test Gold}} & \multicolumn{2}{c|}{\textbf{Test Noisy}} & \multicolumn{2}{c}{\textbf{Test One-shot}} \\
\textbf{Method} & Joint Acc & MRR@3 & Joint Acc & MRR@3 & Joint Acc & MRR@3 \\
\midrule
LAM~\cite{liu2016latent} & 0.24 & 0.27 & 0.23 & 0.26 & 0.12 & 0.19 \\
RecipeGen++~\cite{yusuf2022recipegentpp} & 0.31 & 0.35 & 0.29 & 0.34 & 0.17 & 0.24 \\
TARGE~\cite{cimino2025targe} & 0.58 & 0.63 & 0.39 & 0.44 & 0.45 & 0.49 \\
\midrule
\textbf{FARM (Ours)} & \textbf{0.79} & \textbf{0.84} & \textbf{0.62} & \textbf{0.69} & \textbf{0.70} & \textbf{0.77} \\
\bottomrule
\end{tabular}

\vspace{3mm}  % Add spacing between table and caption
\caption{Comparison with prior trigger-action programming approaches on a
\textbf{service-level} benchmark. We derive service-level labels by mapping each
trigger/action \emph{function} in our dataset to its parent platform service.
We re-ran LAM, RecipeGen++, and TARGE using the authors' released code on this
service-level dataset, using identical train/validation/test splits across all
methods. FARM operates at function granularity, but for this table we map
FARM's predicted functions to their parent services to enable direct comparison.
\textbf{Joint Accuracy} is computed on the top-1 predicted trigger-service and
action-service pair. \textbf{MRR@3} is computed as the mean reciprocal rank of
the ground-truth trigger-action \emph{service pair} within the top-3 ranked
pairs produced by each method.}

\label{tab:baseline_comparison}
\end{table*}

\subsection{Ablation Studies}
\label{subsec:ablation}

We conduct ablation studies to understand the contribution of each system component.

\subsubsection{Layer Freezing Ablation}

\begin{figure}[t]
\centering
\includegraphics[width=0.80\columnwidth]{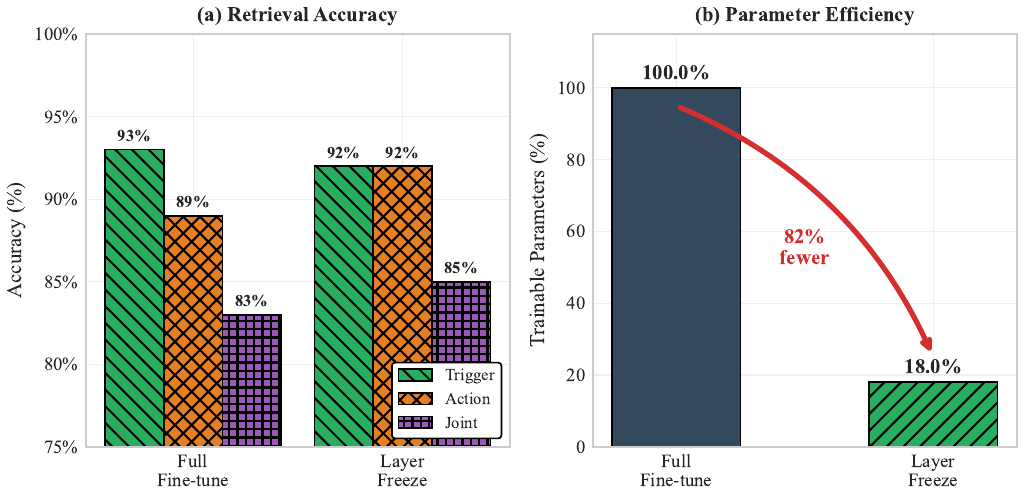}
\caption{Layer freezing ablation using R@5 metric (consistent with Figure~\ref{fig:lora_ablation}). Layer freezing achieves slightly better joint R@5 accuracy (85\%) with 92\%/92\% for trigger/action compared to full fine-tuning (83\% joint) with 93\%/89\% for trigger/action, while using only 18\% of trainable parameters. Note: R@5 is used here instead of R@1 to match the LoRA ablation evaluation protocol, which uses $k$=5 for retrieval evaluation.}
\label{fig:layer_freezing}
\end{figure}

Figure~\ref{fig:layer_freezing} compares full training against layer freezing on retrieval accuracy.

Layer freezing preserves the pretrained encoder's understanding in frozen layers (0--11) while training upper layers (12--23) for domain-specific patterns.

\subsubsection{LoRA vs. Layer Freezing}

\begin{figure}[t]
\centering
\includegraphics[width=0.80\columnwidth]{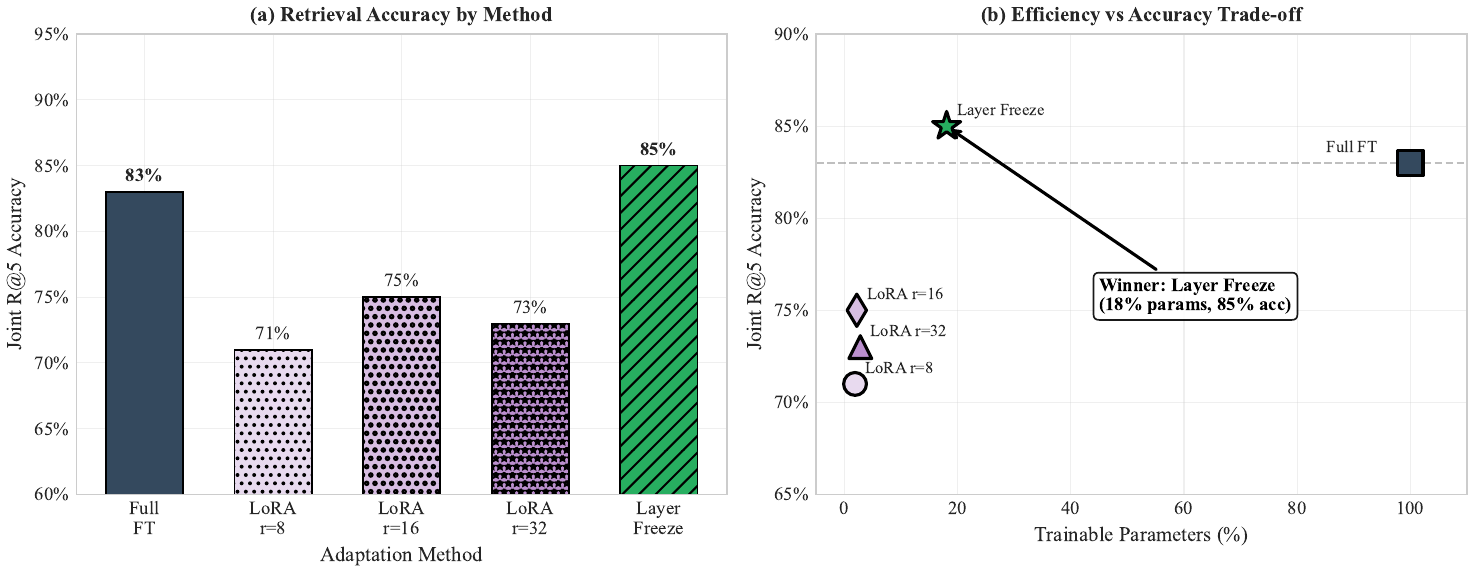}
\caption{LoRA vs. Layer Freezing ablation. (a) Joint R@5 accuracy by adaptation method. Layer freezing achieves the highest accuracy (85\%) while LoRA variants underperform (69--73\%). (b) Efficiency vs. accuracy trade-off showing layer freezing provides the best balance of parameter efficiency (18\%) and retrieval performance.}
\label{fig:lora_ablation}
\end{figure}

We compare our layer freezing approach against Low-Rank Adaptation (LoRA)~\cite{hu2022lora}, the dominant parameter-efficient fine-tuning method. Figure~\ref{fig:lora_ablation} presents results across retrieval accuracy and parameter efficiency.

Layer freezing achieves 85\% Joint R@5 with 18\% trainable parameters, outperforming full fine-tuning (83\% Joint R@5, 100\% params) and all LoRA variants. LoRA's low-rank constraint limits its ability to adapt to the TAP domain, with increasing rank ($r$=8$\rightarrow$32) providing only marginal improvements. Layer freezing provides the best trade-off: highest retrieval performance with moderate parameter cost.

\subsubsection{Component Ablation}

\begin{figure}[!h]
\centering
\includegraphics[width=0.85\columnwidth]{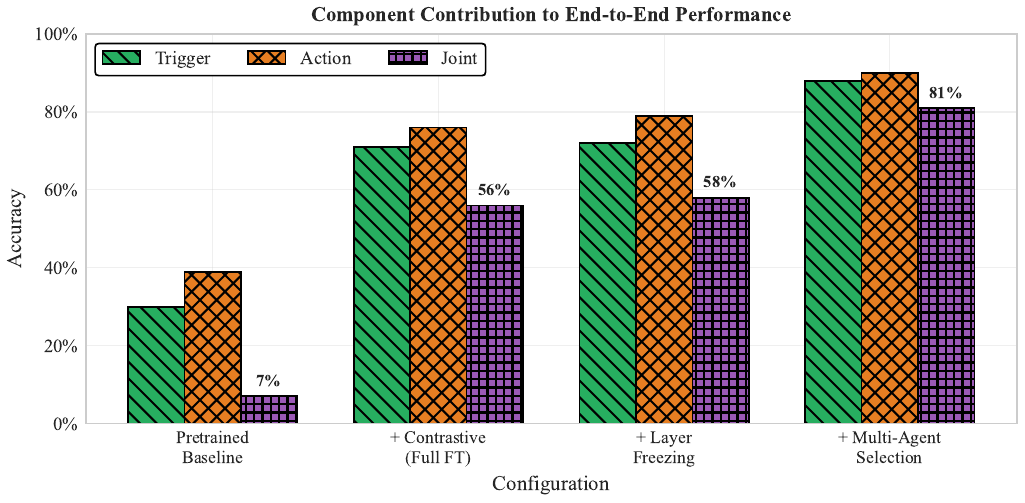}
\caption{Component ablation showing incremental contributions.}
\label{fig:component_ablation}
\end{figure}

\begin{figure}[H]
\centering
\includegraphics[width=\columnwidth]{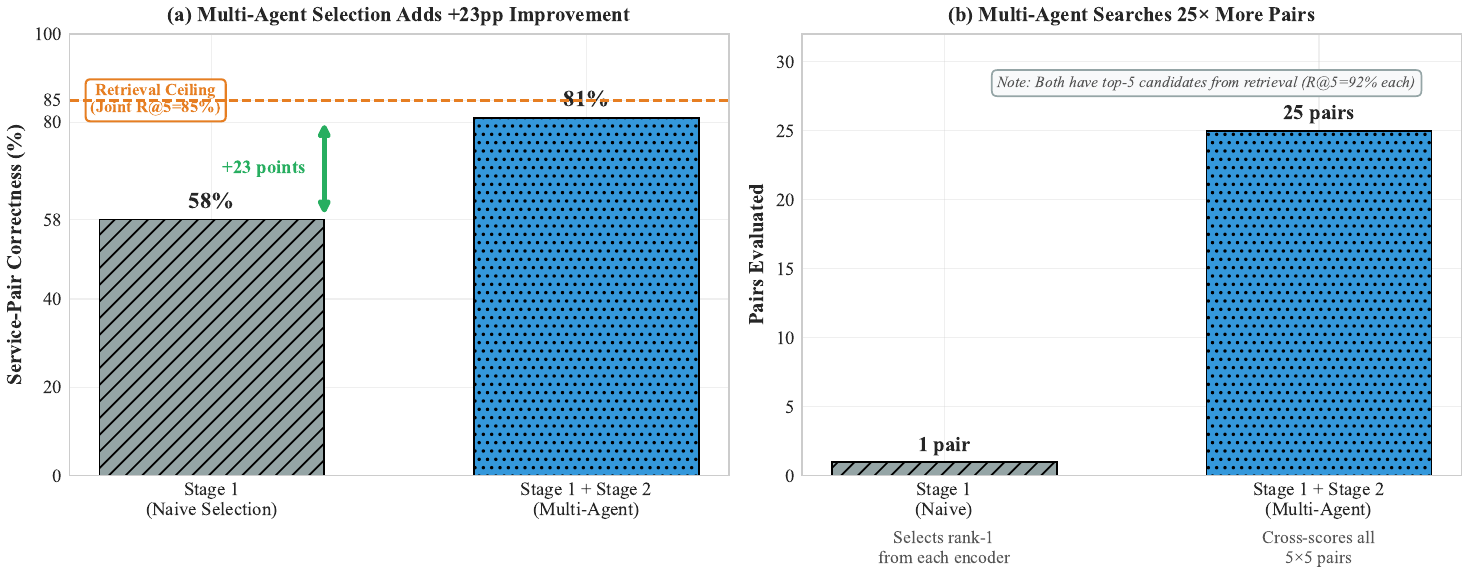}
\caption{Stage 2 improvement mechanism. (a) Multi-agent selection achieves 81\% joint accuracy, a +23 point improvement over naive rank-1 selection (58\%), approaching the 85\% retrieval ceiling (Joint R@5). (b) The improvement stems from evaluating 25 trigger-action pairs (5$\times$5) with LLM-based cross-compatibility scoring, compared to naive selection which evaluates only 1 pair (rank-1 from each encoder).}
\label{fig:stage2_mechanism}
\end{figure}

\noindent\textbf{Component Contributions:}
\begin{enumerate}
    \item \textbf{Contrastive Training (+49 pts Joint R@1):} Improves retrieval top-1 dual success from near-random (7\%) to viable (56\%).
    \item \textbf{Layer Freezing (+2 pts Joint R@1):} Improves retrieval top-1 dual success from 56\% to 58\% while using only 18\% trainable parameters.
    \item \textbf{Multi-Agent Selection (+23 pts Joint Accuracy):} This substantial improvement (58\% $\rightarrow$ 81\%) stems from expanding the search space: naive selection evaluates only 1 pair (rank-1 from each encoder), while multi-agent selection evaluates up to $k^2 = 25$ pairs through cross-compatibility scoring (Fig.~\ref{fig:component_ablation},~\ref{fig:stage2_mechanism}).
\end{enumerate}

% \begin{figure}[H]
% \centering
% \includegraphics[width=\columnwidth]{figures/fig13_stage2_mechanism.pdf}
% \caption{Stage 2 improvement mechanism. (a) Multi-agent selection achieves 81\% joint accuracy, a +23 point improvement over naive rank-1 selection (58\%), approaching the 85\% retrieval ceiling (Joint R@5). (b) The improvement stems from evaluating 25 trigger-action pairs (5$\times$5) with LLM-based cross-compatibility scoring, compared to naive selection which evaluates only 1 pair (rank-1 from each encoder).}
% \label{fig:stage2_mechanism}
% \end{figure}
% \FloatBarrier
\subsubsection{LLM Selection Ablation}

We evaluate the impact of different large language models on the multi-agent selection pipeline. Table~\ref{tab:llm_ablation} presents generation quality metrics for various LLMs suitable for tool calling and agentic workflows.

\begin{table}[t]
\centering
\begin{tabular}{lcc}
\toprule
\textbf{LLM} & \textbf{Faithfulness} & \textbf{Topic Adherence} \\
\midrule
LLaMA 3 70B~\cite{meta_llama3_2024} & 0.38 & 0.74 \\
Qwen2 72B~\cite{qwen2_2024} & 0.41 & 0.76 \\
Mistral Large~\cite{mistral_7b_2023} & 0.36 & 0.71 \\
DeepSeek-Coder~\cite{deepseek_coder_2024} & 0.33 & 0.68 \\
\midrule
\textbf{IBM Granite 4.0 Small}~\cite{ibm_granite4_announcement_2025} & \textbf{0.50} & \textbf{0.82} \\
\bottomrule
\end{tabular}
\caption{LLM Ablation for Multi-Agent Selection. Faithfulness measures factual grounding in retrieved schemas; Topic Adherence measures alignment with automation domains. Higher is better. Note: These metrics represent best-case LLM performance on the Gold set; end-to-end system faithfulness across all datasets (Gold/Noisy/One-Shot) ranges from 0.44--0.48 as shown in Fig.~\ref{fig:quality_metrics}.}
\label{tab:llm_ablation}
\end{table}

IBM Granite 4.0 achieves the highest scores on both metrics, with +9--17 points improvement in Faithfulness and +6--14 points improvement in Topic Adherence compared to alternatives. This performance advantage stems from Granite's optimization for enterprise agentic workflows with native tool calling support, enabling more reliable structured JSON output generation and better instruction following for our agent prompting strategy.

\subsubsection{Base Encoder Selection}

To evaluate the impact of base encoder choice on our contrastive learning framework, we compare performance across four state-of-the-art embedding models with diverse architectures and training objectives. We apply our layer freezing strategy (freezing layers 0--11, training 12--23) to EmbeddingGemma and compare against alternative models to assess both absolute performance and the generalizability of our approach.

EmbeddingGemma with layer freezing achieves the best performance across all datasets and configurations
(58\% Gold, 50\% Noisy, 54\% One-Shot Joint R@1; Trigger R@5 = 92\%, Action R@5 = 92\% on Gold),
validating our ablation findings and demonstrating consistent
robustness to dataset quality variations. Different base encoders respond differently to training
strategies: ModernBERT models benefit from full training (e.g., ModernBERT-large on Gold improves from
50\% to 52\%), whereas BGE and E5 models achieve higher performance under layer freezing, similar to
EmbeddingGemma. This indicates that layer freezing is particularly effective for encoders with strong
multilingual pretraining (EmbeddingGemma, BGE, E5), while architectures optimized for long-context
processing (ModernBERT) benefit from full parameter updates. Performance degradation from Gold to Noisy
datasets follows similar trends across models (7--10 percentage point drops), indicating that dataset
quality affects all architectures comparably. Notably, even under their preferred training strategies,
no alternative model surpasses EmbeddingGemma with layer freezing on any dataset split, confirming both
the effectiveness of our approach and the importance of base encoder selection for trigger-action
retrieval (Table~\ref{tab:embedding_comp}).

\begin{table*}[!h]
\centering
\footnotesize
\begin{tabular}{lccccccccc}
\toprule
& & \multicolumn{4}{c}{\textbf{Layer Freezing (18\%)}} & \multicolumn{4}{c}{\textbf{Full Training (100\%)}} \\
\cmidrule(lr){3-6} \cmidrule(lr){7-10}
& & \multicolumn{3}{c}{\textit{Joint R@1}} & & \multicolumn{3}{c}{\textit{Joint R@1}} & \\
\cmidrule(lr){3-5} \cmidrule(lr){7-9}
\textbf{Base Model} & \textbf{Params} & \textbf{Gold} & \textbf{Noisy} & \textbf{One-Shot} & \textbf{R@5} & \textbf{Gold} & \textbf{Noisy} & \textbf{One-Shot} & \textbf{R@5} \\
\midrule
\textbf{EmbeddingGemma}~\cite{embedding_gemma_2025} & 307M & \textbf{0.58} & \textbf{0.50} & \textbf{0.54} & \textbf{0.92} & \textbf{0.56} & 0.48 & \textbf{0.52} & \textbf{0.91} \\
ModernBERT-base~\cite{modernbert_2024} & 149M & 0.45 & 0.38 & 0.42 & 0.85 & 0.47 & 0.40 & 0.44 & 0.87 \\
ModernBERT-large~\cite{modernbert_2024} & 395M & 0.50 & 0.42 & 0.48 & 0.89 & 0.52 & 0.44 & 0.50 & \textbf{0.91} \\
BGE-base-en-v1.5~\cite{bge_embedding_2023} & 109M & 0.48 & 0.40 & 0.45 & 0.88 & 0.46 & 0.38 & 0.43 & 0.86 \\
E5-large-v2~\cite{e5_embeddings_2022} & 335M & 0.52 & 0.44 & 0.49 & 0.90 & 0.50 & \textbf{0.49} & 0.47 & 0.89 \\
\bottomrule
\end{tabular}
\caption{Base encoder comparison.}
\label{tab:embedding_comp}
\end{table*}
\enlargethispage{2\baselineskip}
\clearpage
%%%%%%%%%%%%%%%%%%%%%%%%%%%%%%%%%%%%%%%%%%%%%%%%%%%%%%%%%%%%%%%%%%%%%%%%%%%%%
\subsection{Efficiency Analysis}
\label{subsec:efficiency}

We analyze FARM's computational efficiency across retrieval, selection, and training phases. Table~\ref{tab:efficiency} summarizes the latency and resource requirements for each system component. All experiments conducted on NVIDIA Tesla V100-SXM2-32GB GPUs (32GB memory).

\begin{table}[H]
\centering
\caption{System Efficiency Metrics}
\label{tab:efficiency}
\begin{tabular}{lr}
\toprule
\textbf{Metric} & \textbf{Value} \\
\midrule
\multicolumn{2}{l}{\textit{Stage 1 (Retrieval)}} \\
Encoding Latency & $\sim$50ms per query \\
Vector Search (Qdrant) & $\sim$5ms for top-5 \\
\midrule
\multicolumn{2}{l}{\textit{Stage 2 (Selection)}} \\
LLM Calls per Query & 2--4 \\
Average Total Latency & $\sim$3s (local inference) \\
LLM Configuration & 3$\times$ V100 GPUs \\
Fallback Rate & $<$5\% \\
\midrule
\multicolumn{2}{l}{\textit{Training}} \\
Encoder Training Time & $\sim$20 minutes \\
Training Hardware & Single V100 GPU \\
Training Data Size & 12,652 pairs \\
\bottomrule
\end{tabular}
\end{table}

The two-stage architecture provides computational efficiency: Stage 1 reduces the search space from 2.2M candidate pairs to $k^2 = 25$ combinations (with $k=5$), enabling Stage 2's LLM reasoning to focus on high-quality candidates. This design is necessary because naive selection (taking rank-1 from each encoder, evaluating only 1 pair) achieves just 58\% joint accuracy, whereas evaluating the full $k \times k$ search space with multi-agent reasoning achieves 81\%. The system uses $k=5$ throughout, matching the R@5 evaluation metrics reported in ablation studies.

%%%%%%%%%%%%%%%%%%%%%%%%%%%%%%%%%%%%%%%%%%%%%%%%%%%%%%%%%%%%%%%%%%%%%%%%%%%%%
\subsection{Discussion}
\label{subsec:discussion}

\subsubsection{Why Two Stages?}

Our two-stage architecture addresses a fundamental trade-off between efficiency and accuracy. Stage 1 provides fast ($\sim$55ms) retrieval that reduces the search space from 2.2 million possible pairs to a manageable candidate set. However, \emph{retrieval ranking alone is insufficient for accurate selection}.

To understand why, consider the selection problem: given top-$k$ candidates from each encoder, we must identify the correct trigger-action pair. A naive approach would take the rank-1 result from each encoder (evaluating only 1 pair), but this achieves just 58\% joint accuracy because it requires \emph{both} encoders to rank their correct items first simultaneously. Individual R@1 scores (Trigger: 72\%, Action: 79\%) compound to approximately 58\% joint success, and when either encoder ranks the correct item at position 2--5, the naive approach fails.

Stage 2 addresses this limitation by evaluating the cross-product of candidates: with $k=5$, the system considers up to $k^2 = 25$ trigger-action pairs (Fig.~\ref{fig:stage2_mechanism}b). Multi-agent selection uses LLM reasoning to score pairs based on schema compatibility, ingredient-to-field matching, and intent alignment---factors invisible to embedding-based retrieval. This expanded search achieves 81\% joint accuracy (Fig.~\ref{fig:stage2_mechanism}a), approaching the retrieval ceiling of 85\% (Joint R@5). The 4-point gap between achieved (81\%) and ceiling (85\%) reflects LLM reasoning errors on queries where correct pairs exist in the candidate set but are incorrectly scored.

Put quantitatively: Stage 2 succeeds on 95\% of retrievable queries (81\% / 85\%), demonstrating that the multi-agent search effectively exploits Stage 1's high-recall candidate sets. Neither stage alone could achieve this: Stage 1 lacks reasoning capabilities, while Stage 2 requires the computational efficiency of Stage 1 to focus LLM evaluation on high-quality candidates.

\subsubsection{Prompt-Based Multi-Agent Design}

Stage 2 operates through \emph{prompting only}---no task-specific fine-tuning. This provides flexibility (agents modified via prompts), interpretability (explicit reasoning traces), and generalization (pretrained LLM knowledge handles edge cases). We mitigate potential output inconsistency through structured schemas and agreement-based override mechanisms.

\subsubsection{Limitations}

\begin{enumerate}
    \item \textbf{Retrieval Ceiling:} At $k$=5, individual recall reaches 92\% and Joint R@5 reaches 85\%. Stage 2 cannot recover if the correct service is not retrieved in Stage 1's top-5, establishing an 85\% theoretical upper bound on joint accuracy.
    \item \textbf{Complex Queries:} Multi-intent queries (e.g., ``turn off lights AND lock door'') require decomposition not yet implemented.
    \item \textbf{Faithfulness:} Moderate scores (0.44--0.48) indicate room for improvement in explicit schema grounding.
    \item \textbf{LLM Dependency:} Stage 2 quality depends on the underlying LLM capability; smaller models may underperform.
\end{enumerate}

Several promising directions emerge from this work:

\begin{enumerate}
    \item \textbf{Execution Validation:} Developing automated evaluation of ingredient-to-field binding correctness through integration with live IFTTT API endpoints. This would enable measuring end-to-end applet execution success rates and identifying failure modes in binding generation.

    \item \textbf{Query Decomposition:} Implementing multi-intent query parsing to handle complex automation requests requiring multiple trigger-action pairs.

    \item \textbf{Reinforcement Learning from Human Feedback:} Training a cross-encoder reranker using Group Relative Policy Optimization (GRPO)~\cite{shao2024deepseekmath}, which eliminates the need for a critic model by estimating baselines from group scores. This would enable learning nuanced scoring functions from user acceptance/rejection signals while remaining memory-efficient. GRPO's compatibility with existing LLM infrastructure makes it particularly suitable for improving Stage 2 selection quality through iterative refinement.

    \item \textbf{Cross-Platform Generalization:} Extending evaluation to other TAP platforms (Zapier, Make, Home Assistant) to validate approach generalizability across different service ecosystems and schema formats.

    \item \textbf{User Studies:} Conducting human evaluation to assess practical utility and user satisfaction with generated applets in real-world scenarios, including longitudinal studies of applet usage patterns.
\end{enumerate}

\section*{Acknowledgment}

This research was supported by the Ministry of Science and ICT (MSIT), Korea, under the Information Technology Research Center (ITRC) program supervised by the Institute for Information and Communications Technology Planning and Evaluation (IITP) (IITP-2025-RS-2023-00259099, 35\%), by the National Research Foundation of Korea (NRF) grant funded by the Korea government (MSIT) (RS-2023-00240211, 35\%), and by the Culture, Sports and Tourism R\&D Program through the Korea Creative Content Agency, funded by the Ministry of Culture, Sports and Tourism in 2024, titled ``Global Expert Personnel Training in MR Space Computing-based Technologies'' (RS-2024-00399186, 30\%).

\bibliographystyle{ACM-Reference-Format}
\bibliography{references}

\appendix

\end{document}